\documentclass[aps,
twoside,nofootinbib,tightenlines,12pt,preprintnumbers,unsortedaddress]{revtex4}
\preprint{YITP-SB-03-64}
\preprint{TUW-03-38}
\preprint{ITP-UH-33/03}
\usepackage{graphicx}
\usepackage{latexsym}
\usepackage{amsmath}
\usepackage{amssymb}
\usepackage{array}
\def\1{\vec } \def\2{\frac{1}{2}} \def\4{\frac{1}{4}}
\def\5{\bar } 
\def\6{\partial }
\def\7#1{{#1}\llap{/}}
\def\8#1{{\textstyle{#1}}} \def\9#1{{\bf{#1}}}
\def\.{\dot }
\def\^#1{\widehat{#1}}
\def\ket#1{|#1\rangle }

  \let\g=\gamma 
   
  \let\l=\lambda \let\m=\mu
    
\let\ph=\varphi

\def\CL{{\cal L}}

\def\({\left(} \def\){\right)} \def\<{\left\langle } \def\>{\right\rangle }
\def\[{\left[} \def\]{\right]}  
 
\def\pmbf#1{\setbox0=\hbox{${#1}$}
        \kern-.025em\copy0\kern-\wd0
        \kern.05em\copy0\kern-\wd0
        \kern-.025em\raise.0433em\box0 }

\def\be{\begin{equation}}
\def\ee{\end{equation}}

\newcommand{\bel}[1]{\begin{equation}\label{#1}}
\def\bea{\begin{eqnarray}}
\newcommand{\beal}[1]{\begin{eqnarray}\label{#1}}
\def\eea{\end{eqnarray}}
\def\nn{\nonumber\\ }

\newcommand{\ds}{\!\not\!\partial}

\newcommand{\intsum}{\sum\!\!\!\!\!\!\!\int}  

\newcommand{\pv}{\phi_{\mathrm V}}

\newcommand{\SUSY}{susy}

\newcommand{\hal}{{\textstyle\frac{1}{2}}}

\newcommand{\vf}{\varphi}
\newcommand{\cl}{\setlength\arraycolsep}
\newcommand{\ve}{\varepsilon}
\newcommand{\fr}[2]{{\textstyle{\frac{#1}{#2}}}}
\newcommand{\ca}[1]{{\cal{#1}}}

\begin{document}

\title{Quantum corrections to mass and central charge
of\\ supersymmetric solitons}
\author{Alfred Scharff Goldhaber} 
\affiliation{
C.~N.~Yang Institute for Theoretical Physics, SUNY Stony Brook, NY 11794-3840 USA}
\author{Anton Rebhan} 
\affiliation{Institute for Theoretical Physics, Vienna University of Technology, Wiedner Hauptstr.~8--10, A-1040, Vienna, Austria}
\author{Peter van Nieuwenhuizen} 
\affiliation{
C.~N.~Yang Institute for Theoretical Physics, SUNY Stony Brook, NY 11794-3840 USA}
\author{Robert Wimmer}
\affiliation{Institute for Theoretical Physics, University of Hannover, Appelstr.~2, D-30167, Germany\\}

     
\begin{abstract}
We review some recent developments in the subject
of quantum corrections to soliton mass and central charge.
We consider in particular
approaches which use local densities for these corrections,
as first discussed by Hidenaga Yamagishi. We then consider dimensional
regularization of the supersymmetric kink in 1+1 dimensions
and an extension of this method to a 2+1-dimensional gauge theory with 
supersymmetric abelian Higgs vortices as the solitons.
\end{abstract}


\maketitle
\newpage
\tableofcontents
\newpage

\section{Introduction: Local densities for quantum energy and central charge}
        A characteristic theme in the work of Hidenaga Yamagishi is the exploration of quantum field contributions to the structure of solitons.  The first effort was his masterful study of the effect on the Dirac electron vacuum of the choice of chiral boundary condition on the electron wave function at the location of a Dirac monopole \cite{Yamagishi:1983wp}.  This problem can be viewed as an appropriate limit of the case of Dirac isospinor fermions interacting with an 't Hooft-Polyakov monopole, in which one member of the isospinor doublet becomes extremely light.  He showed that  there is a fractional electric charge at the monopole  (i.e., dyon charge), in the form of a polarization charge of the electron vacuum.   Thus  the boundary condition in this context expresses the vacuum angle earlier shown by Witten to imply fractional dyon charge in a theory without fermions \cite{Witten:1979ey}.\footnote{Yamagishi's analysis contained as a special case the known fact that in the presence of strictly massless fermions chiral rotation undoes the effect of vacuum angle, so that the latter becomes unobservable.  Also, in light of the main topic of the present work, it is worth noting that Yamagishi computed the (logarithmically divergent) energy associated with the boundary condition.}   
        
        An equally remarkable and original contribution was Yamagishi's introduction, in the context of solitons in one space and one time dimension, of the concept of local quantum energy density \cite{Yamagishi:1984zv}.  He was led to investigate this subject by the fact that different calculations of the quantum energy for a soliton in $\mathcal N=1$ supersymmetric theory gave different results, both zero and nonzero.  He realized that there could be energies associated with the  choice of boundary conditions on the wave functions for field oscillations, governing their behavior at the edges of a large region containing the soliton.  Computing local densities would eliminate this ambiguity, because those contingent boundary energies would be localized far from the soliton center.

        Yamagishi recognized the necessity of proper renormalization of the quantum perturbation theory for obtaining correct results, and described how to do this renormalization and make the resulting calculation systematic.  His conclusion was that there is a nontrivial local energy density in the examples he studied, but that the integrated energy density vanishes.  This conclusion had considerable appeal, because it implemented an identity between the energy and the central charge (whose quantum correction was expected to vanish \cite{Witten:1978mh,Imbimbo:1984nq}).   Indeed Yamagishi calculated also the local central charge density, verifying the equality of the two integrated densities.  The aforementioned equality arises because in such theories, of the two supersymmetry generators (corresponding to the two components of a fermion wave function), one remains unbroken even in the presence of the soliton.   The difference between energy and central charge is proportional to the norm of a state generated by action of a supersymmetry operator on the soliton ground state.  Thus, if one such operator is unbroken, i.e.,  annihilates the ground state, the equality holds.

        Well over a decade passed after this work, during which there was little further development towards consensus on the correct value of the quantum mass correction for solitons in such 2D supersymmetric theories.  The next stage began when Rebhan and van Nieuwenhuizen  [RV] realized
that a naive energy cutoff used in much of the susy kink
literature was inconsistent with other, more carefully defined regularization
methods. They used mode number regularization (described below) to compute  the total quantum correction for a supersymmetric kink with periodic boundary conditions on the fermion as well as boson fluctuation wave functions  \cite{Rebhan:1997iv}.\footnote{They made this choice of boundary conditions following the original work for a kink in purely bosonic theory by Dashen, Hasslacher, and Neveu [DHN] \cite {Dashen:1974cj}.  Amusingly, in the bosonic theory, the choice of boundary conditions makes no difference to the result.}   Of course, this nonzero result, differing from some prior results, was open to the objection that it might include a boundary energy contribution.   Not long after, the same two authors  joined by Nastase and Stephanov [NRSV] imposed `natural' or `invisible' boundary conditions on the fermion wave functions both in the trivial or vacuum background and in the kink background (so that there could be no extra energy localized near the boundary), and reproduced one of the finite answers from the earlier period \cite{Nastase:1998sy}.  Graham and Jaffe [GJ] did a calculation in terms of scattering phase shifts, with no explicit dependence on boundary conditions, and obtained the same result \cite{Graham:1998qq}.

        While this new activity agreed with the result first found by Schonfeld in his consideration of a kink-antikink system with the two objects well separated \cite{Schonfeld:1979hg}, a complete understanding required reconciliation of the global result with the local calculation of Yamagishi.  Shifman, Vainshtein, and Voloshin [SVV]  \cite{Shifman:1998zy}, stimulated by the work of the Stony Brook--Vienna group, decided to tackle the question head-on by returning to the calculation of the local central charge density
pioneered by Yamagishi.  They identified an ambiguity hidden in the earlier work.  Yamagishi had been explicit in his computation of renormalization effects, but had not discussed explicitly the other `r' required for a one loop calculation:  regularization.   The point here, which Yamagishi clearly understood, is that in calculating an energy sum one must subtract from the sum in the presence of the soliton the equally (quadratically) divergent sum in a flat, or vacuum background.\footnote{Actually in the supersymmetric system the quadratic divergence in the trivial sector cancels between bosons and fermions, but this just replaces the previous regulation problem by one of correctly matching the energy sums for the bosons and fermions in the kink sector.}  The subtracted quantity is only logarithmically divergent, and that is the divergence compensated by renormalization.  Given the regularization scheme, one still has to fix the finite parts in the renormalization by suitable renormalization conditions.  Lacking an explicit regularization scheme (possibly he had simple energy cutoff in mind), Yamagishi was not in a position to determine uniquely the finite part of the renormalization.  By implementing higher-derivative regularization for the central charge density, SVV were able to accomplish this while explicitly maintaining supersymmetry. 
They obtained a nonzero integrated correction to the central charge, precisely equal to the energy shift previously calculated by  Schonfeld, as well as NRSV and GJ.

        Thus the equality of mass and central charge which Yamagishi had found 
was preserved, but both quantities were shifted from zero by an extra 
finite piece which was not expected.  NRSV,  on the basis of their 
nonzero result for the quantum correction to the energy, suspected the 
existence of an anomaly, given the naive expectation that the quantum 
correction to the central charge should vanish.   GJ did not discuss 
anomalies at all, but did notice that the terms corresponding to each 
state in the unregulated, unrenormalized energy-density sum are equal 
one to one with the corresponding terms in the central charge density.  
They accomplished regulation and renormalization for energy by 
subtracting from the divergent sum the divergent contribution from the 
first Born  approximation to the phase shift,  and in the case of 
fermions also the divergent part of the second Born approximation. They 
justified this choice by noting that the no-tadpole condition in the 
trivial sector requires a counterterm for the soliton mass which is 
precisely equal to the integral over momentum of the contribution from  
the subtracted divergent phase shifts.  
  They took the momentum cutoff used to define the counterterm equal to 
the momentum cutoff in the sum over momenta of contributions from 
continuum states to the energy shift.  Having already determined the 
counterterm (using naive energy cutoff), by this choice they implicitly 
defined a regularization scheme.
  To implement the equality of central charge with 
energy, they used the identical subtraction to define  central charge 
as they had used for energy.  This certainly was consistent, but it 
left open the question whether and if so where an anomalous 
contribution to the central charge might be identified, as is remarked 
at the end of the paper of SVV.         In any case, what remained to be 
demonstrated was that one could deduce the regularization and 
renormalization of the central charge directly by analyzing its 
field-theoretic structure.  SVV did exactly that, showing explicitly 
that there is an anomaly, which preserves supersymmetry by providing a 
shift in the central charge equal to that in the energy.  Like the 
famous chiral anomaly this one is an automatic consequence of using a 
regularization scheme that preserves a relevant symmetry, in the 
present case supersymmetry.

        In an important respect the line of development begun by Yamagishi still was incomplete, because the local energy density corresponding to the established global energy and local central charge density was not computed directly.  That was accomplished by Goldhaber, Litvintsev, and van Nieuwenhuizen [GLV2] \cite{Goldhaber:2001rp} and independently by Wimmer \cite{Wimmer:2001yn}.  They demonstrated by explicit checks the validity of a local regularization scheme  called {\it local mode regularization}. This scheme makes computation of energy density of a soliton in 2D theory very clear and simple.   Not surprisingly, it confirms directly for the supersymmetric kink the equality of energy density and central charge density proposed by Yamagishi and demonstrated by SVV.  At the same time, local energy density is an interesting quantity in non-supersymmetric theories, where the definition of central charge is not clear.

        Even with the above results, there remains an open frontier in the computation of quantum energy densities, especially in the direction of higher dimensions.  Clearly, the tree of local energy determination which Hidenaga Yamagishi planted continues to grow and bear fruit, a tribute to his insight and originality.

In section \ref{sec:kink} we focus on a development which was essential to the new understanding of local energy and central charge densities, namely, the careful regularization by the older, global methods of the total mass of a soliton, as discussed in GLV1 \cite{Goldhaber:2000ab}.  This discussion is followed by an explicit demonstration of the universal equivalence, in results for the anomalous or high-energy contribution to quantum energy of a field in the presence of some background potential, between higher-derivative regularization and local mode regularization.

In section \ref{sec:sk} and \ref{sec:vor}, 
we describe some recent work in (slightly) higher dimensions.
We begin by embedding the supersymmetric kink as a domain wall in
2+1 dimensions and use this to set up a supersymmetry-preserving
dimensional regularization scheme. This allows us to derive the
anomaly in the central charge in a particularly transparent manner,
which is found to be made possible by parity violation
and a corresponding nonvanishing expectation
value for the momentum operator in the extra dimension.
In section \ref{sec:vor} we apply this method to
the supersymmetric abelian vortex 2+1 dimension and
determine its quantum corrections to mass and central charge.
\section{Solitons in 1+1 dimensions}
\label{sec:kink}

\subsection{Fermion zero modes in global mode regularization}
One method of regularization, utilized by DHN in their investigation of the quantum correction to the mass of a kink or a sine-Gordon soliton in a purely bosonic theory, is called {\it mode number regularization}.  There is a very concrete picture behind this scheme.  Imagine introducing, in what initially is vacuum, some background potential which influences the motion of a fermion obeying the Schr\"{o}dinger equation.  Suppose further that at the start one had a free Fermi sea filled to some level, so that all possible states up to a maximum number $N$ were occupied.  Then, provided the boundary conditions on the wave functions  at large positive and negative values of the spatial coordinate $x$ prevent any probability leakage through the boundaries, the total number $N$ will not change when the potential is introduced. Further, if the Fermi level is high enough compared to the magnitude of the potential, then there will be no level crossings near that level as the potential is gradually altered.  Thus, just as before the potential was introduced, the first $N$ states will be filled, though the maximum energy may be changed slightly, and the low-energy spectrum may be altered dramatically, for example, to include bound states.    To find the regulated quantum shift in energy one may subtract the sums of energies of the first $N$ states with and without the potential present.  The net sum might still be divergent with $N$, in which case renormalization would be required as well, but in any case the effect of this regularization would be well defined.  

One now can adopt the same maneuver for a relativistic Bose field in the presence of a potential.  Before the potential is introduced, one uses an energy regularization to say that all solutions of the wave equation up to a maximum energy $\epsilon_{\rm max}$ are counted.  There are $N$  pairs of these solutions, each pair with one positive and one (equal) negative frequency.  After introducing the potential, we again require $N$ pairs of solutions, and those solutions are all the ones having energy $\epsilon_{i}^{\prime}\le \epsilon_{\rm max}^{\prime}(N)$.  Thus we find a regulated, but not yet renormalized, energy,
 \begin{equation}
 E_{\rm reg}=\frac12\sum_{i=1}^{N}(\epsilon^{\prime}_{\rm i} -\epsilon_{\rm i})  \   \  , \label{one}
 \end{equation}
  where the factor $\frac12$ is the familiar coefficient in the zero-point energy contribution of each mode.

A subtlety arises in this calculation if the presence of a potential happens to introduce a zero-frequency solution, which would occur either accidentally or as a result of some particular symmetry obeyed by the potential.  For example, if the potential for Bose fluctuations arises from the  presence of a nontrivial classical soliton solution of the equations of motion of the Bose field, then there is clearly a zero frequency deformation corresponding to uniform translation of the soliton.  As mentioned above, solutions of (equal) positive and negative frequency come in pairs, and one pair defines one state. At zero frequency there is also one state,  because the `momentum' conjugate to the Bose field for the translation solution becomes the momentum of the soliton moving with some definite velocity.    Each pair of solutions of the bosonic field equations (including any at zero frequency) corresponds to one term in (\ref{one}).

For fermions the situation is less familiar.  Again, there are paired positive and negative frequency solutions, each corresponding to a term in  (\ref{one}).  However, if there is only one zero energy solution of the fermionic fluctuation equations, this does not correspond to a term in  (\ref{one}).  The coefficient of this solution in the expansion of the Fermi field is an operator $c_{0}$ satisfying $c_{0}^{2}=1$, and that $c_{0}$ appears in the expression for one of the two supersymmetry operators (the one that does not annihilate the soliton ground state).  Several remarkable facts follow from these observations.  First, such a wave function does not correspond to a zero mode of the soliton, because any mode requires two independent, noncommuting nilpotent creation and annihilation operators, and here one has instead one hermitean, idempotent operator (which may be represented as a Pauli $\sigma$ matrix).     This immediately introduces a serious difficulty for the standard recipe of mode regularization, because the spectrum of nonnegative frequency solutions no longer is in one-to-one correspondence with the number of terms in (\ref{one}).   Secondly, because the one supersymmetry operator associated with the zero-frequency solution does not annihilate the soliton ground state, it follows that the ground state spontaneously breaks supersymmetry.   

There is another remarkable aspect of this phenomenon.  The ground state not only is not annihilated by a supersymmetry operator, but becomes an eigenstate of that operator, if one decomposes the kink vacua $|K\rangle$ and $c_{0}|K\rangle$ into  $\frac{1}{\sqrt{2}}(1+c_{0})|K\rangle$ and  $\frac{1}{\sqrt{2}}(1-c_{0})|K\rangle$ \cite{Ritz:2000xa}.  This is remarkable because there is a discrete $Z_{2}$ operator which takes the Fermi field into its negative, and which in other contexts could be used to prove that one cannot build a coherent superposition of a boson and a fermion.  Now we find that an operator anticommuting with this $Z_{2}$ operator leaves the ground state unchanged.  Therefore   two states with equal and opposite eigenvalues for the supersymmetry generator must be identified. In other words, the Hilbert space splits into two noncommunicating parts, which are gauge copies of each other \cite{Goldhaber:2000ab}.  This is a discrete analogue to the Higgs mechanism, in which a continuous local gauge symmetry is hidden because a scalar field in a nontrivial representation of that gauge symmetry has a nonzero expectation value.

Something not considered in GLV1 is the generalization of this discussion about the  $Z_{2}$ gauge symmetry to the case of 
many solitons, which would be relevant for sine-Gordon solitons if not for the kink (the focus of that study).\footnote{Because a kink cannot be adjacent to another kink, pure multi-kink systems, with no anti-kinks, are not possible.} 
The $Z_{2}$ symmetry is not a local gauge symmetry, so that for a multisoliton system the {\it relative} signs of eigenvalues of the supersymmetry operator acting on different solitons {\it might} be significant.  This may be related to a discovery of Moore and Read long before, in a 2+1-dimensional system which appears to describe the 5/2 state of the fractional quantum Hall effect \cite{Moore:1991ks}.  They discussed `nonabelian statistics' for objects carrying electric charge $\mp e/4$.  This kind of statistics seems to us interpretable in terms of the eigenvalues of the supersymmetry operator, which would reverse sign when one such object made a full circle around another.   Specifically, for a system with $n$ solitons separated from each other by a finite distance, charge conjugation symmetry guarantees $[n/2]$ pairs of equal positive and negative frequency solutions, and for odd $n$ one zero frequency solution.  To each of the paired solutions corresponds a fermion creation operator, and hence there are $2^{[n/2]}$ nearly degenerate states of the n-soliton system.  For infinite separation, there are $n$ operators $c_{0}$ which each have equal positive and negative eigenvalues.  Superficially this might imply $2^{n}$ distinct states, but as we already have argued there is a gauge redundancy which reduces this number.  The calculation for finite separation shows that the reduction must bring that number down to $2^{[n/2]}$.

While these ideas are surprising and beautiful, one would like to have a clear idea what has happened to the principle of mode number regularization, now that the counting of fermion modes suddenly has a lacuna.  GLV1 found two different ways to deal with this question.   First, following NRSV, one may use locally invisible (periodic or antiperiodic) boundary conditions in vacuum, and locally invisible (twisted periodic (TP) or twisted antiperiodic (TAP), where the twist refers to a chiral rotation of a wave function by $\pi/2$ at the one boundary with respect to that at the other) conditions in the presence of a kink.  Now an assumption necessary to derive the equality of numbers of modes before and after the background is introduced no longer holds, because the twisting at the boundary allows `leakage' of mode number at the boundary.  This is a concept that goes back to  an article of Goldstone and Wilczek describing adiabatic flow of fermion number out of a magnetic monopole as the Yukawa coupling of the fermions to a Higgs field is chirally rotated \cite{Goldstone:1981kk}.  In their discussion the leakage is out to infinity rather than into the boundary.  The final result for localized  charge is the same, whether the chiral rotation is applied to the mass, while the boundary condition is fixed, or applied to the boundary condition, while the mass remains constant (as in Yamagishi's work on monopole dyon charge \cite{Yamagishi:1983wp}).   The leakage in our example involves exactly half a unit of mode number, and this shift reconciles the count of fermion modes with the principle of mode number regularization.  

A second approach is to use fixed boundary conditions, which insures no leakage of mode number, but to average over a set of boundary conditions, so that in the average any contributions to the energy localized near the boundary cancel out.  In this case, the
mode number regularization is used for each set of fixed boundary conditions, but in some cases there are two zero-frequency solutions in the kink sector, one localized near the kink and one localized near the boundary, and in some cases there is one in the kink sector localized at the kink and one in the vacuum sector localized near the boundary.  When the difference between the sectors is two solutions in the kink sector, this implies one (delocalized) zero mode in the kink sector.  Thus the average over all boundary conditions amounts to half a zero mode excess in the kink sector.  

We see that the two different approaches agree:  Effectively there is a half-mode at zero energy in the kink sector. This was discussed already by GJ
(who analyzed half-modes at zero energy as well as 
nonlocalized `half-bound' modes at the continuum threshold). 
What GLV1 added was a precise specification of the boundary conditions which make this notion well defined.  
The half-mode has the consequence for the mode sums used to compute the energy that at the cutoff one needs to include an excess of one half-mode in the sum for the vacuum energy compared to the sum for the kink energy.  One may recognize this fact directly by matching modes so as to insure that no difference between the two sums linear in the energy will occur.  Thus, as with many other examples of anomalies and associated regulation prescriptions, one may find the correct prescription and its implications either by focusing on the lowest energies (in this case zero modes), or on the highest energies. 

The half mode bears an interesting relation to the original discovery by Jackiw and Rebbi \cite{Jackiw:1976fn} that a soliton can polarize a charged fermion vacuum to localize a half unit of fermion number around the soliton.  For neutral excitations, there is no directly observable charge or charge density that corresponds to a mode or to the probability density in its wave function.  Nevertheless modes of nonzero energy contribute to vacuum zero point energy and energy density.  Thus mode number and mode density are mathematically well defined and indirectly observable, making them also meaningful, even if not so tangible as the directly observable charge and charge density of a charged field. In other words, mode number 1/2 has operational significance just as does charge 1/2.
A further indication of this significance is additivity:  Two configurations each possessing a zero frequency solution corresponding to mode number 1/2, when brought within a finite distance of each other, accommodate one observable excitation with near-zero energy.  Thus, even though it is not possible to define directly a ``half-excitation'' it is possible to count a half-mode, and to combine two of them to make not only a full mode but also a full excitation or fermion state.

Although the invisible boundary conditions guarantee that the quantum correction is precisely the mass shift of the soliton, there remains even in this case a strong reason for averaging over each of the two types of condition, namely to enforce time-reversal symmetry.  A chiral or twisted boundary condition produces different spectra for right-moving and left-moving  waves.  Time reversal symmetry interchanges TP and TAP boundary conditions, so that by averaging over the two one restores the symmetry.  In the vacuum or trivial sector, periodic and antiperiodic conditions (being real) each separately produce spectra invariant under the symmetry.  

\subsection{Local mode regularization from higher-derivative regularization}
SVV in their discussion emphasized the value of using a local regularization scheme which enforces supersymmetry, in particular, higher-derivative regularization.  In GLV2 it was shown that a familiar scheme, point--splitting regularization, implies local mode regularization for energy density.  Here we show directly that higher-derivative regularization also implies local mode regularization for energy density, thus verifying explicitly its consistency with supersymmetry.  

In higher derivative regularization,  extra terms are added to the Lagrangian   involving extra factors of the square of the spatial gradient of the field.  This gives rise to an equation for small fluctuations of a Bose field, including a possible background `potential' $V(x)$,\footnote{For simplicity we focus here only on the pure Bose field case, but this time there is no difficulty in generalizing the procedure to include Fermi fields as well.}
\begin{equation}
 \omega^{2}(1-\partial_{x}^{2}/\Lambda^{2})\phi=-\partial_{x}^{2}\phi +m^{2}\phi +V(x)\phi + \partial_{x}^{4}\phi /\Lambda^{2} \  \  . \label{hdr}
\end{equation}
Our notation uses  $\Lambda$ as the regulator mass called $M_{r}$ by SVV. The effect of the higher-derivative term on the classical kink solution and the resulting potential $V(x)$ for fluctuations  tends to zero for large $\Lambda$, and therefore is ignored in our discussion, i.e., we use the $V$ obtained from the classical solution of the unregulated equations. 
Here 
as in GLV2, we may use the JWKB approximation to estimate the wave functions at large $\omega$.  This allows one to determine with sufficient accuracy  the phase space density of these wave functions both for the trivial or vacuum background with $V=0$ and the kink background with 
\begin{equation}
V=-3m^{2}/2\cosh^{2}(mx/2) \ \ ,
\end{equation}
 where $m$ is the mass for small fluctuations about the trivial vacuum.
Then one may obtain a subtracted density which can be integrated to give the high-energy contribution to the regulated net local mode density, as we shall now see.

Concentrate on a high-energy regime, with wave vectors $|k|>K\gg \sqrt{V}$, with $\omega^{2}\equiv k^{2}+m^{2}/(1+k^{2}/\Lambda^{2})$, and take $\Lambda \gg K$.  
Write the wave function as $\phi(x)=e^{ikx}e^{if(x)}$, and make a gradient expansion,
$f(x)=f^{(0)}(x)+f^{(1)}(x)+.... \  \ $:
\begin{equation}
2kf^{(0)\prime}(1+\frac{k^{2}}{\Lambda^{2}})=-V(x) \ \ .
\end{equation} 
and 
$$2kf^{(1)\prime}(1+\frac{k^{2}}{\Lambda^{2}})=-\partial_{x}^{2}if^{(0)}(1+5\frac{k^{2}}{\Lambda^{2}})=$$
\begin{equation}
-i\partial_{x}V(x)(1+5\frac{k^{2}}{\Lambda^{2}})\bigg{/}2k(1+\frac{k^{2}}{\Lambda^{2}})  \  \  .
\end{equation}
Because $f^{(1)}$ is imaginary, it alters the modulus of $\phi$, and because $ f^{(1)}$ and $V$ both vanish at spatial infinity, this equality relating first derivatives of these two functions implies an equality relating the functions themselves.

We need now to compute the mode density difference between the case with and without nonzero V(x).  Define normalization of the wave function $\phi_{k}(x)$ so that $\phi^{2}$ is equal to unity far from the region of the potential. The regulated mode density before subtraction is then 
\begin{equation}\
d\rho (x) /d|k| =\phi_{k}(x)^{*}\frac{1}{1-\partial_{x}^{2}/\Lambda^{2}}\phi_{k}(x)/\pi  \ \ .\label{bubby}
\end{equation} 
To lowest order in $V/K^{2}$ and $ V/\Lambda^{2}$ we now compute the mode density difference 
\begin{equation}
d\delta\rho /d|k|=\frac {V(x)}{2\pi}\bigg{\{}\frac{1}{k^{2}}\bigg{[}5\frac{k^{2}}{\Lambda^{2}}+1\bigg{]} + \frac{1}{\Lambda^{2}}\bigg{[}\frac{k^{2}}{\Lambda^{2}}+1\bigg{]} \bigg{\}}\bigg{/}\bigg{[}\frac {k^{2}}{\Lambda^{2}}+1\bigg{]}^{3} \ \  . \label{brace}
\end{equation}
Here the first term in the large braces comes from  the influence of $V(x)$ on the absolute square of the wave functions, while the second term comes from the influence of $V$ on the regulating denominator in Eq. ({\ref{bubby}).  To first order in $V$, this denominator factor is
\begin{equation}
[1-\partial_{x}^{2}/\Lambda^{2}]^{-1}\approx\bigg{\{}1+\frac{k^{2}}{\Lambda^{2}} -\frac{V(x)}{\Lambda^{2}}\bigg{/}\bigg{[}1+\frac{k^{2}}{\Lambda^{2}}\bigg{]}\bigg{\}}^{-1} \ \  .
\end{equation}
Omitting terms of order $1/\Lambda$, the integral of (\ref{brace}) over the range $|k|=K$ to $|k|=\infty$  gives the extra contribution at high energies to the local mode density (number of modes per unit length)
\begin{equation}
\rho_{\rm mode}(x)=V(x)/2\pi K \  \  .
\end{equation}
Note that this result is insensitive to $\Lambda\gg K$. 

   In the original local mode regularization approach, this extra density was treated as all associated with momentum $|k|=K$, so that the resulting `anomalous' extra energy density resulting from the regularization is 
\begin{equation}
{\cal E}_{\rm anom}(x)=\frac K2 \rho_{\rm mode}(x)=V(x)/4\pi  \  \ ,
\end{equation}
\\
including a correction by a factor -2 to Eq.(57) of GLV2, 
Now, for the full calculation with higher-derivative 
regularization, we must compute $\int_{|k|=K}^{|k|=\infty}d|k| (d\delta\rho /dk) (|k|/2)$, where the factor $|k|/2$ at the end is the zero-point energy corresponding to a given mode (neglecting quantities of ${\cal O}( m^{2}/k^{2})$).  This integral is easily evaluated, yielding a total 
high-energy contribution to the regulated (but not yet renormalized) energy density
\begin{equation}
{\cal E}_{\rm unrenormalized}(x)=\frac{V(x)}{8\pi}\bigg{[}2+\ell n\frac{ \Lambda^{2}} {K^{2}}\bigg{]} \ \ . \label{2}
\end{equation}
Here the anomaly term 2 in Eq. (\ref{2}) bears exactly the same ratio to the term (compensated by mass renormalization) $\ell n \Lambda^{2}$ as found by SVV in their analysis of the supersymmetric kink, e.g., their Eq. (4.5).  This ratio of course is trivial in local mode regularization, as is easy to see from the description above.  However, in Eq. (\ref{2}), the term 2 comes half from the first term in Eq. (\ref{brace}) and half from the second,
while the logarithm clearly comes entirely from the first.

Looking at this straightforward but still somewhat intricate computation, one may understand why SVV focused on calculation of the anomaly in the central charge density, and did not  attack directly the anomaly in the energy density.  However,  Eq. (\ref{2}) shows that the result of local mode regularization for the anomaly is universally correct in higher-derivative regularization, and thus completes a circle:  

DHN used (global) mode regularization for the mass  of the bosonic kink, but refrained from tackling the problem with fermions. Yamagishi introduced local energy and central charge density calculations, but did not fully address the role of regularization in these calculations.  A variety of approaches using global methods managed to respect constraints of regulation and still circumvent the difficulty of contributions to quantum energy from boundary conditions on fermions.

  SVV made a complete calculation of the central charge density (both for the anomaly and for other contributions) but did not compute energy density directly, so that their method was not applicable in its original form  for purely bosonic structures (or indeed any structure without supersymmetry).
GLV2 introduced an efficient method, local mode regularization, for computing the anomaly in the energy density of any system, but previous history of 
regularization schemes implies that one should check whether a new scheme preserves symmetries which we want keep at the quantum level, in this case supersymmetry.  The present calculation, verifying local mode regularization for arbitrary background potentials starting from higher-derivative regularization (which does preserve supersymmetry), brings us back to the beginning with a reliable local scheme for doing what DHN did globally. 

There is another interesting aspect of local mode regularization.  The derivation from point-splitting regularization in GLV2 involves a shift in the real part of the effective wave number at a given value of $k$.  However, the original method, and also the computation done here, involves an integration over $k$ of the imaginary part of the wave number.  This has a suggestive resemblance to dispersion relations connecting the real part of a scattering amplitude to an integral over the imaginary part, related by unitarity to a cross section.

\subsection{Why is rigorous, manifest supersymmetry unnecessary for correct calculation?}
We have seen that a completely rigorous and systematic way to assure consistency with supersymmetry in computing the quantum correction to the mass and central charge of a  soliton is to insist on explicit supersymmetry at every step.  Thus SVV adopted the approach of higher-derivative regularization in the ultraviolet and supersymmetric boundary conditions discretizing the spectrum to provide infrared regularization.  They then computed a local central charge density in the vicinity of the kink, and its spatial integral in a large but finite region around the soliton, which gives the desired quantum correction.   While this is the safest method, and also fairly straightforward because the central charge density has a weaker superficial divergence than the energy density, a number of other methods all yield the same result.  

It seems to us that this not only demonstrates that the result is correct, but also suggests that there is a deeper principle behind the robustness of that result.  The principle, we suggest, is cluster decomposition.  In GLV1 it was deduced erroneously from an incomplete calculation that besides localized energy in the vicinity of the kink and possible localized energy near the boundary, there can be a delocalized energy, giving a finite contribution.  They imposed the principle of cluster decomposition to sum over different boundary conditions, so that the delocalized energy violating cluster decomposition would cancel out.   It was observed by Wimmer \cite{Wimmer:2001yn}, and argued in  detail in the light of the discrete C,P. and T symmetries by Goldhaber, Rebhan, van Nieuwenhuizen, and Wimmer \cite{Goldhaber:2002mx} that such delocalized energy does not appear in mode number regularization for known  boundary conditions.  Cluster decomposition is satisfied automatically.  Thus, a boundary condition which violates the supersymmetry preserved by the kink can alter the energy localized near the boundary, but cannot influence the energy localized near the soliton, i.e., the quantum correction to the kink mass.   This means that if one uses boundary conditions which do not result in boundary energy, or if one averages over conditions so that there is no net boundary energy, then the resulting global calculation is reliable for the mass of the kink.  This is true even if the boundary conditions do not preserve supersymmetry -- that violation does not propagate into the region of the kink.

As pointed out in GLV2, even methods whose consistency with supersymmetry has not been explicitly established, such as point-splitting regularization, give the same answer not only for the energy density but also for the central charge density.  Thus  these local densities are quite robust with respect to the choice of  (local) regularization methods.  This suggests that, despite the long history of discrepant calculations, one must make quite a large error to get the wrong answer for the local densities.  For example, failure to specify carefully the regularization method can make the result indeterminate, and therefore most likely wrong.  The history of incorrect global calculations draws attention in particular to regulation by energy cutoff, which was used in many of those calculations.  As local mode regularization makes manifest, using a fixed and identical energy cutoff for kink and trivial backgrounds is  incorrect.  Litvintsev and van Nieuwenhuizen \cite{Litvintsev:2000is} proposed a method to repair energy cutoff, which is mathematically equivalent to replacing it with mode number regularization.

\section{Dimensional regularization of the susy kink}
\label{sec:sk}

Usually dimensional regularization breaks susy. However,
the 1+1 dimensional $\mathcal N=1$
susy kink can be embedded as a domain wall in
2+1 dimensions with the same field content while keeping
$\mathcal N=1$ susy invariance.

For the corresponding classically
BPS saturated domain wall (a 1+1 dimensional object by itself), 
\cite{Rebhan:2002uk} has also found a 
nontrivial quantum correction to the energy
density. 
In order to have BPS saturation at the quantum level
in 2+1 dimensions, there
has to be a matching correction to the momentum in the extra
dimension which corresponds to the central charge of the 1+1
dimensional case.

In this section we show that if one uses 
susy-preserving dimensional regularization
by means of dimensional reduction from 2+1 dimensions, 
one indeed finds the required correction
to the extra momentum. Such a nonvanishing correction turns out to
be possible because the 2+1 dimensional theory spontaneously breaks
parity.

By dimensionally reducing to 
1+1 dimensions, the parity-violating contributions
to the extra momentum turn out to provide an
anomalous contribution to the central charge as obtained
in Ref.~\cite{Shifman:1998zy}, thereby giving a novel physical
explanation of the latter \cite{Rebhan:2002yw}. 
This is in line with the well-known fact that central charges of
susy theories can be reinterpreted as "momenta" in higher dimensions.

The latter statement has to be handled with care, though.
The classical
central charge stems entirely from the \emph{classical antisymmetric} part 
of the energy momentum tensor of the $2+1$ dimensional theory and
thus would be missed by dimensional reduction 
if one were to start 
in $2+1$ dimensions with the gravitational energy momentum tensor,
which is always symmetric on-shell (in the absence of local
Lorentz anomalies) and 
which contains the {genuine} momentum operator. However,
it is
the \emph{symmetric}
part of the $2+1$ dimensional EM-tensor which gives the anomalous 
contribution to the \emph{quantum} central charge. This anomalous contribution
can be reduced to a surface term and is
thus completely determined by the topology of the soliton background,
independent of the precise field profile in the bulk. 
Therefore when we refer to the $\vf^4$ kink in the following, this is 
just a special case of a more general situation.

In the case of the susy kink, standard ('t Hooft-Veltman)
dimensional regularization is seen
to be compatible with susy invariance only at the expense of a 
spontaneous parity
violation, which in turn allows non-vanishing quantum corrections
to the extra momentum in one higher spatial dimension.
On the other hand, as we shall see, the surface term that
provides the classical central charge does not receive quantum 
corrections in dimensional regularization, by the same reason
that led to null results previously in other schemes
\cite{Imbimbo:1984nq,Rebhan:1997iv,Nastase:1998sy}.
In dimensional regularization (by going up in the number of dimensions),
the nontrivial anomalous quantum correction to the central charge operator
is thus seen to be entirely the remnant of the spontaneous parity violation
in the higher-dimensional theory in which a susy kink
can be embedded by preserving minimal susy.

Alternatively, we shall consider dimensional regularization
by dimensional reduction from 1 to 1-$\epsilon$ spatial
dimensions, which also preserves supersymmetry.
In this case we show that an anomalous contribution to the central
charge arises from the necessity to add an evanescent
counterterm to the susy current \cite{Rebhan:2002yw}. 
This counterterm preserves susy but
produces an anomaly in the conformal-susy current. 
We also construct the conformal central-charge current \cite{Rebhan:2002yw}
whose
divergence is proportional to the
ordinary central-charge current and thus
contains the central-charge anomaly as
superpartner of the conformal-susy anomaly \cite{Shifman:1998zy}.

\subsection{The model}

A real scalar field model in 1+1 dimensions with 
spontaneously broken $Z_2$ symmetry ($\vf\rightarrow-\vf$) has topologically
nontrivial finite-energy solutions called ``kinks'' which
interpolate between the two neighboring degenerate vacuum states,
as for example  $\vf=\pm v$.
If the potential is of the form $V(\vf)=\hal U^2(\vf)$ it has a minimal%
, $\ca{N}=(1,1)$, supersymmetric extension 
\cite{DiVecchia:1977bs}
\bel{Lssan}
\CL=-\2\left[ (\6_\m\vf)^2+U(\vf)^2+\5\psi\g^\m\6_\m\psi+
U'(\vf)\5\psi\psi \right]
\end{equation}
where $\psi$ is a Majorana spinor, $\5\psi=\psi^{\mathrm T} C$ with
$C\gamma^\mu=-(\gamma^\mu)^TC$. We use
a Majorana representation of the Dirac matrices with $\g^0=-i\sigma^2$,
$\g^1=\sigma^3$, and $C=\sigma^2$ in terms of the
standard Pauli matrices $\sigma^k$ so that $\psi=\binom{\psi^+}{\psi^-}$ with
real $\psi^+(x,t)$ and $\psi^-(x,t)$.
(The reason for choosing $\g^1=\sigma^3$, rather than $\g^1=\sigma^1$, is
that it diagonalizes the Dirac equation.)

The $\vf^4$ model 
is defined as the special case
\begin{equation}\label{Uphian}
U(\vf)=\sqrt{\frac{\l}{2}}\(\vf^2-v_0^2\),\qquad v_0^2\equiv \mu_0^2/\l
\end{equation}
where the $Z_2$ symmetry of the susy action  
also involves the fermions
according to $\vf\to-\vf, \psi\to\gamma^5\psi$ with 
$\gamma^5=\gamma^0\gamma^1$.
A classical kink at rest at $x=0$ which interpolates
between the two vacua is given by \cite{Rajaraman:1982is}
\bel{Ksol}
\ph_{K}=v_0 
\tanh\(\mu_0 x/\sqrt2\).
\ee%

At the quantum level we have to renormalize, and we shall employ
the simplest possible scheme\footnote{See \cite{Rebhan:2002uk} for
a detailed discussion of more general renormalization schemes
in this context.}
which consists of putting all
renormalization constants to unity except for a mass counter term
chosen such that tadpole diagrams cancel completely in
the trivial vacuum. 
So we set $Z_\varphi=Z_\psi=Z_\lambda=1$ and
$\mu_0^2=\mu^2+\delta \mu^2$, for which 
at the one-loop
level and using dimensional regularization we find
\begin{equation}\label{deltamu2}
\delta \mu^2 = \lambda\, \delta v^2 = \lambda \int\frac{dk_0d^{d}k}{(2\pi)^{d+1}}\frac{-i}{k^2+m^2-i\epsilon}= 
\lambda\int\frac{d^{d}k}{(2\pi)^d}\frac{1}{2[\vec k^2+m^2]^{1/2}},
\end{equation}
where $m=U'(v)=\sqrt{2}\mu$ is the tree-level mass of elementary bosons and
fermions, and $k^2=\vec k^2-k_0^2$. 

The susy invariance of the model (\ref{Lssan})
under 
\begin{equation}
  \label{eq:anz1}
  \delta\varphi=\bar\epsilon\psi\ \ ,\ \ 
  \delta\psi=(\not\!\partial\vf-U)\epsilon,
\end{equation}
(with $\mu_0^2$ replaced by
$\mu^2+\delta\mu^2$) leads to the on-shell conserved Noether
current
\begin{equation}\label{susyj}
j_\mu=-(\not\!\partial\vf+U(\vf))\gamma_\mu\psi
\end{equation}
and two conserved charges $Q^\pm=\int dx\,j^{0\pm}$.

The model (\ref{Lssan}) is equally supersymmetric in 2+1 dimensions,
where we use $\gamma^2=\sigma^1$, and the Noether current and
charges are unchanged in form. The same renormalization scheme
can be used,
only the renormalization constant (\ref{deltamu2}) has
to be evaluated for $d=2-\epsilon$ in place of $d=1-\epsilon$
spatial dimensions.

While classical kinks in 1+1 dimensions 
have finite energy (rest mass) $M=m^3/\lambda$,
in 
2+1 dimensions 
they yield domain walls with
a profile given by (\ref{Ksol}) and
finite surface (string) tension
$M/L=m^3/\lambda$. With
a compact extra dimension one can use these configurations
to form ``domain strings'' of finite total energy proportional
to the length $L$ of the string when
wrapped around the extra dimension. 

The 2+1 dimensional case is different also with respect to the
discrete symmetries of (\ref{Lssan}). In 2+1 dimensions, 
$\gamma^5=\gamma^0\gamma^1\gamma^2 = \pm {\bf 1}$ corresponding
to the two inequivalent  irreducible representations for $\gamma^2=\pm\sigma^1$.
Therefore, the sign of the fermion mass (Yukawa) term can no longer
be reversed by $\psi\to\gamma^5\psi$ and there is no longer
the $Z_2$ symmetry  $\vf\to-\vf, \psi\to\gamma^5\psi$.

What the 2+1 dimensional model does break spontaneously instead is
{\em parity}, which corresponds
to changing the sign of one of the spatial coordinates.
The Lagrangian is invariant under $x^m \to -x^m$ for
a given spatial index $m=1$ or $m=2$  together with $\vf\to-\vf$ (which
thus is a pseudoscalar) and $\psi\to\gamma^m \psi$.
Each of the trivial vacua breaks these invariances spontaneously,
whereas a kink background in the $x^1$-direction with
$\vf_K(-x^1)=-\vf_K(x^1)$ preserves
$x^1=x$ reflection symmetry, but breaks it with respect to $x^2=y$.
 
This is to be contrasted with the 1+1 dimensional case, where
parity ($x^1\to-x^1$) can be represented either by $\psi\to\gamma^0\psi$
and a true scalar $\vf\to\vf$ or by $\psi\to\gamma^1\psi$ and
a pseudoscalar $\vf\to-\vf$. The former leaves the trivial vacua
invariant, and the latter the ground state of the kink sector.

\subsection{Susy algebra}

The susy algebra for the 1+1 and the 2+1 dimensional cases can
both be covered by starting from 2+1 dimensions, the 1+1 dimensional
case following from reduction by one spatial dimension.

In 2+1 dimensions one obtains \cite{Gibbons:1999np}
\begin{eqnarray}
  \label{eq:3dsusy}
 \{Q^{\alpha},\bar{Q}_{\beta}\}&=&2i(\gamma^M)^{\alpha}{}_{\beta}P_M\ ,
          \quad (M=0,1,2)\nonumber\\
    &=&2i(\gamma^0H+\gamma^1(\tilde{P}_x+\tilde{Z_y})
    +\gamma^2(\tilde{P}_y-\tilde{Z}_x))^\alpha{}_\beta,
\end{eqnarray}%
where we separated off two surface terms $\tilde Z_m$ in defining
\begin{eqnarray}
  \label{eq:Ptilde}
\tilde P_m = \int d^dx  \tilde{\mathcal{P}}_m, \quad 
  &&\tilde{\mathcal{P}}_m=\dot\vf\,\partial_m\vf
      -\2(\bar{\psi}\gamma^0\partial_m\psi),\\
  \label{eq:Ztilde}
\tilde Z_m = \int  d^dx  \tilde{\mathcal{Z}}_m, \quad
  &&\tilde{\mathcal{Z}}_m=U(\vf) \partial_m\vf = \partial_m \ca{W}(\vf)
\end{eqnarray}
with $\ca{W}(\vf)\equiv\int d\vf\, U(\vf)$. Note that the usual central 
charge density of the two-dimensional model, $\tilde{\mathcal{Z}}_m$, 
is obtained by dimensional reduction of the antisymmetric part of the 
three-dimensional energy momentum tensor. The local version of the susy 
algebra (\ref{eq:3dsusy}) is obtained by a susy variation (\ref{eq:anz1})
of the supercurrent (\ref{susyj}) as follows
{\cl{2pt}
\begin{eqnarray}
  \label{eq:anz2}
   T^{MN}&\sim&Tr(\gamma^M \delta j^N)=Tr(\gamma^M\gamma^N\gamma^P)
   \ \partial_P\vf U(\vf)+\mathrm{symm.\ part}\nonumber\\
   &\sim&\ve^{MNP}\partial_P\vf U(\vf) +\mathrm{symm.\ part},
\end{eqnarray}}%
and the central charge density is then the momentum density  $T^{02}$ in the 
reduced extra dimension.

With our choice of Dirac matrices the supercharges 
and the superalgebra they generate are given by 
\begin{eqnarray}
  \label{eq:qpm}
  &&Q^{\pm}=\int d^2x[(\dot\vf\mp\partial_y\vf)\psi^\pm
      +(\partial_x\vf\pm U(\vf))\psi^\mp],\\ \label{qpm2}
&&\{Q^\pm,Q^\pm\}=2(H \pm (\tilde Z_x - \tilde P_y)),\ \ 
\{Q^+,Q^-\}=2(\tilde P_x+\tilde Z_y)\ .
\end{eqnarray}
Having a kink profile in the $x$-direction, which satisfies the
Bogomol'nyi equation $\partial_x \vf_K=-U(\vf_K)$, one finds that 
the charge $Q^+$ (corresponding to the terms in (\ref{eq:anz1})
with $\epsilon^-$)
leaves the classical topological (domain-wall) vacuum $(\vf=\vf_K,\ \psi=0)$
invariant.
This corresponds to classical
BPS saturation, since with $P_x=0$ and $\tilde P_y=0$
one has $\{Q^+,Q^+\}=2(H+\tilde Z_x)$ and, indeed, with a kink domain wall
$\tilde Z_x/L^{d-1}={\cal{W}}(+v)-{\cal{W}}(-v)=-M/L^{d-1}$.

At the quantum level, hermiticity of $Q^\pm$ and positivity of the 
Hilbert space norm imply a lower bound for the energy(density):
\begin{equation}
  \label{HPyineq}
   \< \Sigma|H|\Sigma\> \ge |\< \Sigma|P_y|\Sigma\>|  
   \equiv |\langle \Sigma|(\tilde P_y-\tilde Z_x)|\Sigma\rangle|,
\end{equation}
where $|\Sigma\rangle$ denotes any state in the Hilbert space.
This inequality is saturated when
\begin{equation}
  \label{eq:anzc2}
 Q^+|\Sigma\rangle=0.
\end{equation}
Massive BPS states in 1+1 dimensions correspond to massless states
in 2+1 dimensions, since with $[H,P_m]=0$  one has 
\begin{equation}
  \label{eq:anzc1}
  \langle P_M P^M\rangle=-\fr{1}{4}\langle(Q^{+2}Q^{-2}-\{Q^+,Q^-\}^2\rangle=0 
\end{equation}
for BPS saturated states (\ref{eq:anzc2}) with 
$\langle P_y\rangle=M$ for a kink domain wall with
kink profile in the $x$-direction.
An anti-kink domain wall has instead $ Q^-|\Sigma\rangle=0$. In both cases,
half of the supersymmetry is spontaneously broken.
To take into account that 
there is  infinite momentum and energy unless the $y$-direction
is compact with finite length $L$, one can formulate the above identities
for energy and central charge per unit length or for
energy and central charge densities.

Omitting  regularization  the susy algebra in 1+1 dimensions is obtained from
(\ref{eq:3dsusy}) simply by dropping $\tilde P_y$
as well as $\tilde Z_y$ so that $P_x\equiv\tilde P_x$.
The term $\gamma^2 \tilde Z_x$ remains, however, with
$\gamma^2$ being the nontrivial $\gamma^5$ of 1+1 dimensions.
Identifying $\tilde Z_x$ with $Z$,
the susy algebra simplifies to
\begin{equation}
\{Q^\pm,Q^\pm\} = 2(H\pm Z),\quad \{Q^+,Q^-\}=2P_x
\end{equation}
and one obtains the quantum BPS bound  
\begin{equation}
\langle\psi|H|\psi\rangle \ge |\langle\psi|Z|\psi\rangle|
\end{equation}
for any state $| \psi\rangle$. BPS saturated states have
$Q^+|\psi\rangle=0$ or $Q^-|\psi\rangle=0$, corresponding to
kink and anti-kink, respectively, and break half of the
supersymmetry.


\subsection{Fluctuations}

In a kink (or kink domain wall) background one spatial
direction is singled out and we choose this to be along $x$.
The direction orthogonal to the kink direction (parallel
to the domain wall) will be denoted by $y$.

The quantum fields can then be expanded in the
eigenfunctions, which are known analytically for the
$\vf^4$ and sine-Gordon soliton \cite{Rajaraman:1982is}, times plane waves in
the extra dimensions. For the bosonic fluctuations we have
\be
[-\square+(U{'}^2+UU{''})]\eta=0
\ee 
which is solved by
\begin{equation}
  \label{eq:bosfluct}
\eta=\int\frac{d^{d-1}\ell}{(2\pi)^{\frac{d-1}{2}}}
\intsum\frac{dk}{\sqrt{4\pi\omega}}
    \left(a_{k,\ell}\ e^{-i(\omega t-\ell y)}\phi_k(x)+
      a^{\dagger}_{k,\ell}\ e^{i(\omega t-\ell y)}\phi^{\ast}_k(x)\right).
\end{equation}
The kink eigenfunctions $\phi_k$ are 
normalized according to $\int dx|\phi|^2=1$ for the discrete states 
and to Dirac distributions for the continuum states
according to $\int dx\, \phi_k^*\phi_{k'}=2\pi\delta(k-k')$.
The latter are
deformed plane waves because there is no reflection in the case of the kink.
The mode energies are 
$\omega=\sqrt{\omega_k^2+\ell^2}$
where $\omega_k$ is the energy in the 1+1-dimensional case.   

The canonical equal-time commutation relations 
$[\eta(\vec{x}),\dot{\eta}(\vec{x}')]=i\delta(\vec{x}-\vec{x}')$
are fulfilled with
\begin{equation}
  \label{eq:ETC}
  [a_{k,\ell},a^{\dagger}_{k',\ell'}]=\delta_{kk'}\delta(\ell-\ell'),
\end{equation}%
where for the continuum states $\delta_{k,k'}$ becomes a Dirac delta
function. 

For the fermionic modes which satisfy the Dirac equation $[\ \ds+U{'}]\psi=0$,
i.e.\ explicitly
\bea
&&(\partial_x+U')\psi^++i(\omega+\ell)\psi^- = 0, 
\nonumber\\
&&(\partial_x-U')\psi^-+i(\omega-\ell)\psi^+ = 0\label{psipm},
\nonumber
\eea
one finds
\begin{eqnarray}
  \label{eq:ferm}
 \psi\!&=&\!\psi_0+\int\frac{d^{d-1}\ell}{(2\pi)^{\frac{d-1}{2}}}
\intsum'\frac{dk}{\sqrt{4\pi\omega}}
 \left[b_{k,\ell}\  e^{-i(\omega t-\ell y)}
      \binom{{ \scriptstyle\sqrt{\omega+\ell}}\ \phi_k(x)}{
                    {\scriptstyle\sqrt{\omega-\ell}}\ is_k(x)}
            + b^{\dagger}_{k,\ell}\ (c.c.)\right],\nonumber\\
 &&\psi_0=\int\frac{d^{d-1}\ell}{(2\pi)^{\frac{d-1}{2}}}b_{0,\ell}\ e^{-i\ell (t- y)}
   \binom{\phi_0}{ 0},\quad b^{\dagger}_0(\ell)=b_0(-\ell).
\end{eqnarray}
The fermionic zero 
mode\footnote{By a slight abuse of notation
we shall always label this by a subscript $0$, but this should not be
confused with the threshold mode $k=0$ (which does not appear
explicitly anywhere below).}
of the susy kink turns into
massless modes located on the domain wall, which have only one
chirality, forming a Majorana-Weyl domain wall fermion 
\cite{Rebhan:2002uk,Callan:1985sa,Gibbons:2000hg,Hofmann:2000pf}.\footnote{The 
mode with $\ell=0$
corresponds in 1+1 dimensions to the fermionic
zero mode of the susy kink. If there are no other fermionic zero modes,
it has to be counted as half a degree of freedom in mode
regularization \cite{Goldhaber:2000ab}. For dimensional
regularization such subtleties do not play a role because
the zero mode only gives scale-less integrals and these vanish.}

For the massive modes the Dirac equation relates the eigenfunctions
appearing in
the upper and the lower components of the spinors as follows:
\begin{equation}
  \label{eq:sin}
  s_k=\frac{1}{\omega_k}(\partial_x+U{'})\phi_k
     =\frac{1}{\sqrt{\omega^2-\ell^2}}(\partial_x+U{'})\phi_k,
\end{equation}%
so that the function $s_k$ is the 
SUSY-quantum mechanical \cite{Witten:1982df} partner of 
$\phi_k$ and thus coincides with the eigen modes of the 
sine-Gordon model if $\phi_k$ belongs to the $\vf^4$-kink 
(hence the notation) \cite{Cooper:2001zd}.
With (\ref{eq:sin}), their normalization is the same as that of
the $\phi_k$. It is the relation (\ref{eq:sin}) and
the fact that it relates bosonic to fermionic modes, as well
as different components of the fermionic modes to each other,
which makes it possible to compute the one loop-corrections to 
energy and central charge 
without explicit knowledge of the mode functions.

The canonical equal-time anti-commutation relations 
$\{\psi^{\alpha}(\vec{x}),\psi^{\beta}(\vec{x}')\}=\delta^{\alpha\beta}
\delta(\vec{x}-\vec{x}')$ 
are satisfied if (using that $\phi_{-k}(x)=\phi_k^*(x)$ and
$s_{-k}(x)=s_k^*(x)$)
\begin{eqnarray}
  \label{eq:fermetc}
  \{b_0(\ell),b^{\dagger}_0(\ell')\}&=&\{b_0(\ell),b_0(-\ell')\}=\delta(\ell-\ell'),
   \nonumber\\
  \{b_{k,\ell},b^{\dagger}_{k',\ell'}\}&=&\delta_{k,k'}\delta(\ell-\ell'),
\end{eqnarray}%
and again the $\delta_{k,k'}$ becomes a Dirac delta for the continuum states. 
The 
algebra (\ref{eq:fermetc}) 
and the solution for the massless mode (\ref{eq:ferm})
show that the operator $b_0(\ell)$  creates right-moving
massless states on the wall when $\ell$ is negative and 
 annihilates them for positive  momentum $\ell$. 
Thus only massless states with 
momentum in  the positive $y$-direction can be created. 
Changing the representation 
of the gamma matrices by $\gamma^2\rightarrow-\gamma^2$, 
which is inequivalent to 
the original one, reverses the situation. 
Now only massless states with momenta in 
the negative $y$-direction exist. Thus depending on the representation
of the Clifford algebra one chirality of the domain wall fermions
is singled out. This is a reflection of the spontaneous violation 
of parity when embedding the susy kink as a domain wall in 2+1 dimensions.

Notice that in (\ref{eq:ferm}) $d$ can be only 2 or 1, for which
$\ell$ has 1 or 0 components, so for strictly $d=1$ one has $\ell\equiv0$. 
In order to have a susy-preserving dimensional regularization
scheme by dimensional reduction, we shall start from $d=2$ spatial
dimensions, and then make $d$ continuous and smaller than 2.

\subsection{Energy corrections}

Before turning to a direct calculation of the anomalous
contributions to central charge and momentum, we derive
the one-loop  energy density of the
susy kink (domain wall) in dimensional regularization.

Expanding the Hamiltonian density of the model (\ref{Lssan}),
\begin{equation}
\label{eq:ham}
  \mathcal{H}=\2[\dot{\vf}+(\nabla\vf)^2+{U}^2(\vf)]
   +\2 \psi^{\dagger}i\gamma^0[\vec{\gamma}\ \vec{\nabla}+U'(\vf)]\psi,
\end{equation}%
around the kink/domain wall, using $\vf=\vf_K+\eta$, one obtains
\begin{eqnarray}
  \label{eq:hamexp}
  \mathcal{H}&=&\2[(\partial_x\vf_K)^2+{U}^2]
  -\frac{\delta\mu^2}{\sqrt{2\lambda}}U-\partial_x(U\eta)+{}\nonumber\\
  &&+\2[\dot{\eta}^2+(\vec\nabla\eta)^2+\2(U^2){''}\eta^2]
      +\2\psi^{\dagger}i\gamma^0[\vec{\gamma}\ \vec{\nabla}+U{'}]\psi+
O(\hbar^2),
\end{eqnarray}
where $U$ without an explicit argument implies evaluation at $\vf=\vf_K$
and use of the renormalized $\mu^2$.
The first two terms on the r.h.s. are the classical energy density
and the counterterm contribution. The terms quadratic in the
fluctuations are the only ones contributing to the
total energy.\footnote{The third term in (\ref{eq:hamexp}) is of relevance
when calculating the energy profile \cite{Shifman:1998zy,Goldhaber:2001rp}.} 
They give \cite{Rebhan:2002yw}
\begin{eqnarray}
  \label{eq:h22}
   \langle\mathcal{H}^{(2)}\rangle&=&-\partial_x
 \left(\2\int\frac{d^{d-1}\ell}{(2\pi)^{d-1}}\intsum \frac{dk}{2\pi} \ 
   U{'}\frac{|\phi_k|^2}{2\omega}\right){}\nonumber\\
 &&{}+\2\int\frac{d^{d-1}\ell}{(2\pi)^{d-1}}\intsum \frac{dk}{2\pi}
   \frac{\ell^2}{2\omega}(|\phi_k^2|-|s_k|^2).
\end{eqnarray}%
When integrated, the first term, which is a pure surface term,
cancels exactly the counterterm (see (\ref{deltamu2})), because
\begin{eqnarray}
  \label{eq:surface}
  \int dx\langle\2\partial_x(U{'}\eta^2)\rangle=
    \2 U{'}\langle\eta^2\rangle|_{-\infty}^{\infty}=
    m\int\frac{d^{d-1}\ell}{(2\pi)^{d-1}}\int\frac{dk}{2\pi}\frac{1}{2\omega}
\equiv m \delta v^2,
\end{eqnarray}%
where we have used that $U{'}(x=\pm\infty)=\pm m$.

In these expressions, the massless modes (which correspond to the zero mode
of the 1+1 dimensional kink) can be dropped in dimensional regularization
as scale-less and thus vanishing contributions, and the massive discrete
modes cancel between bosons and fermions.\footnote{The zero mode
contributions in fact do not cancel by themselves
between bosons and fermions, because the
latter are chiral. This non-cancellation is in fact crucial in energy cutoff
regularization (see Ref.~\cite{Rebhan:2002uk}).}
Using the
explicit form of $\phi_k(x)$, $s_k(x)$ as given e.g.\ in
\cite{Goldhaber:2001rp},
the $x$-integration over the continuous mode functions
gives the required difference of spectral densities as
\begin{equation}\label{specdiff}
\int dx (|\phi_k(x)|^2-|s_k(x)|^2)=-\theta'(k)
=-\frac{2m}{k^2+m^2},
\end{equation}
where $\theta(k)$ is the additional phase shift of the mode functions
$s_k$ compared to $\phi_k$. 

With the help of the susy-quantum mechanical relation (\ref{eq:sin}) for
the fermionic modes in the BPS background the integral 
(\ref{specdiff}) can also be computed without detailed knowledge of the mode 
functions \cite{Wimmer:2003PhD}. Denoting the operator in (\ref{eq:sin}) by 
$A=\partial_x+U{'}$ the fluctuation equation above (\ref{eq:bosfluct})
of the "bosonic" modes $\phi_k$ factorizes as 
\begin{equation}
  \label{eq:anz5}
  A^\dagger A\phi_k=\omega^2_k\phi_k,
\end{equation}%
where $A^\dagger=-\partial_x+U{'}$ is the adjoint operator. 
Using  (\ref{eq:sin}) the spectral density 
(\ref{specdiff}) can be written as
{\cl{2pt}
\begin{eqnarray}
  \label{eq:anz6}
  -\theta'(k)&=&\int dx\ [|\phi_k(x)|^2-\frac{1}{\omega_k^2}(A\phi_k)^\ast(A\phi_k)]
  \nonumber\\&=&\int dx\ [|\phi_k(x)|^2-\frac{1}{\omega_k^2}\phi_k^\ast 
  A^\dagger A\phi_k]
             +\textrm{surface term}.
\end{eqnarray}}
The first term simply vanishes because of (\ref{eq:anz5}). The surface term
results from the fact that in the above expression
the operator $A^\dagger$ is 
only formally
the hermitian 
conjugate of $A$.
The difference of the spectral densities is therefore given
by
\begin{equation}
  \label{eq:anz7}
   -\theta'(k)=-\frac{1}{\omega_k^2}\int dx\ \partial_x(U{'}|\phi_k(x)|^2)
          =-\frac{2m}{k^2+m^2},
\end{equation}%
where we have omitted a term $\phi^* \partial_x \phi$
because it is
is even in $x$ for large $|x|$
and thus only needed the asymptotic values 
$U{'}[\vf_K(x=\pm\infty)]=U{'}(\pm v)=\pm m$ and that the mode functions 
are plane waves asymptotically, i.e. $|\phi_k|^2\rightarrow 1$. This result
coincides with the one from
a direct evaluation of (\ref{specdiff}).

Thus we obtain for the remaining correction from (\ref{eq:h22})
\begin{equation}
 \label{Mkink}
  \frac{\Delta M^{(1)}}{L^{d-1}}
  =-\4\int \frac{dk\,d^{d-1}\ell}{(2\pi)^d}\frac{\ell^2}{\omega}\theta'(k)
  =-\frac{2}{d}\frac{\Gamma(\frac{3-d}{2})}{(4\pi)^{\frac{d+1}{2}}}\, m^d.
\end{equation}
This reproduces the correct, known result for the susy kink mass
correction $\Delta M^{(1)}=-m/(2\pi)$ (for $d=1$) and
the surface (string) tension of the
2+1 dimensional susy kink domain wall 
$\Delta M^{(1)}/L=-m^2/(8\pi)$ (for $d=2$) 
\cite{Rebhan:2002uk}.

Notice that the entire result is produced by an integrand
proportional to the extra momentum component $\ell^2$,
which for strictly $d=1$ would not exist.

\subsection{Anomalous contributions to the central charge and
extra momentum}
\label{sec33}

In a kink (domain wall) background with only nontrivial $x$ dependence,
the central charge density $\tilde{\mathcal{Z}}_x$ receives
nontrivial contributions.
Expanding $\tilde{\mathcal{Z}}_x$ 
around the kink background gives
\begin{eqnarray}
  \label{eq:Znaiv}
  \tilde{\mathcal{Z}}_x=U\partial_x\vf_K-\frac{\delta\mu^2}{\sqrt{2\lambda}}
        \partial_x\vf_K+\partial_x(U\eta)+\2\partial_x(U{'}\eta^2)
        +O(\eta^3). 
\end{eqnarray}
Again only the part quadratic in the fluctuations contributes to
the integrated quantity at one-loop order\footnote{Again, this does not hold
for the central charge density locally 
\cite{Shifman:1998zy,Goldhaber:2001rp}.}.
However, this leads just to the contribution shown in (\ref{eq:surface}),
which matches
precisely the counterterm $m\delta v^2$ from requiring vanishing tadpoles.
Straightforward application of the rules of
dimensional regularization thus leads to a null result for
the net one-loop correction to $\langle\tilde Z_x\rangle$ in the same way
as found in Refs.~\cite{Imbimbo:1984nq,Rebhan:1997iv,Nastase:1998sy}
in other schemes.

On the other hand, by considering the less singular
combination $\langle H+\tilde Z_x\rangle$ and showing that it vanishes exactly, 
it was concluded in Ref.~\cite{Graham:1998qq}
that $\langle\tilde Z_x\rangle $ has to compensate any nontrivial result
for $\langle H\rangle$, which in Ref.~\cite{Graham:1998qq} was obtained
by subtracting successive Born approximations for scattering
phase shifts. In fact, Ref.~\cite{Graham:1998qq} explicitly
demonstrates how to rewrite $\langle\tilde Z_x\rangle$ into 
$-\langle H\rangle$,
apparently without the need for the anomalous terms in the quantum central
charge operator
postulated in Ref.~\cite{Shifman:1998zy}.

Because the authors of Ref.~\cite{Graham:1998qq}
did not discuss regularization of
$\langle\tilde Z_x\rangle$, the manipulations
needed to rewrite it as $-\langle H\rangle$ (which eventually is
regularized and renormalized) are not defined in their work. 
If we choose to use dimensional regularization, $\langle\tilde Z_x\rangle$
contains the mode energies $\omega=\sqrt{k^2+m^2+\ell^2}$ instead of  
$\omega_k=\sqrt{k^2+m^2}$ and so the manipulations carried through in  
Ref.~\cite{Graham:1998qq} (eq. 56) are no longer possible.
Using dimensional regularization one in fact obtains
a nonzero result for $\langle H+\tilde Z_x\rangle$, apparently
in violation of susy.
However, dimensional regularization by embedding the kink
as a domain wall in (up to) one higher dimension, which
preserves susy, instead leads to 
\begin{equation}
\langle H+\tilde Z_x-\tilde P_y\rangle =0,
\end{equation}
i.e. the saturation of (\ref{HPyineq}), as we shall now verify.

The bosonic contribution to $\langle\tilde P_y\rangle$ involves an $\ell$ integral
which is scale-less and odd and thus vanishes. 
Only the fermions turn out to give interesting contributions:
\begin{eqnarray}
  \label{eq:py}
  \langle\tilde{\mathcal{P}}_y\rangle&=&
  \frac{i}{2}\langle\psi^{\dagger}\partial_y\psi\rangle\nonumber\\
  &=&\2\int\frac{d^{d-1}\ell}{(2\pi)^{d-1}}\intsum' \frac{dk}{2\pi}
     \left( \frac{\ell}{2}(|\phi_k|^2+|s_k|^2)
    +\frac{\ell^2}{2\omega}(|\phi_k|^2-|s_k|^2)
\right).\qquad
\end{eqnarray}
We have already omitted the contributions which vanish either by symmetric integration
or due to scale-less integrals, which are zero in dimensional regularization. The remaining 
$\ell$-integration no longer
factorizes because $\omega=\sqrt{k^2+\ell^2+m^2}$, and
is in fact identical to the finite contribution
in $\<\mathcal H\>$ obtained already in (\ref{eq:h22}):
\begin{equation}
  \label{eq:anz8}
  -\Delta Z=\int dx\ \langle\tilde{\mathcal{P}}_y\rangle=
  \4\int \frac{dk\,d^{d-1}\ell}{(2\pi)^d}\frac{\ell^2}{\omega}\theta'(k)
  =\frac{2}{d}\frac{\Gamma(\frac{3-d}{2})}{(4\pi)^{\frac{d+1}{2}}}\, m^d.
\end{equation}%

So for all $d\le2$ we have BPS saturation, $\langle H\rangle=|\langle\tilde Z_x-\tilde P_y\rangle |$,
which in the limit $d\to1$, the susy kink, is made possible by 
a non-vanishing $\langle\tilde P_y\rangle$. The anomaly in the central charge
is seen to arise from a parity-violating contribution in $d=1+\epsilon$
dimensions which is the price to be paid for preserving supersymmetry
when going up in dimensions to embed the susy kink as a domain wall.

It is again the difference in the spectral densities, $\theta{'}$, which
determines the one-loop corrections, which  thus depend 
only on the derivative of the
pre-potential (or equivalently the second derivative of 
super potential $\ca{W}=\int d\vf U(\vf)$) 
at the critical points $\pm v$. 
In general the spectral density difference 
for a model with spontaneously broken  $\mathbb{Z}_2$ symmetry is given
by 
\begin{equation}
  \label{eq:anz9}
  \theta{'}(k)=\frac{\ca{W}{''}(v)-\ca{W}{''}(-v)}{k^2+\ca{W}{''}(v)^2},
\end{equation}
which has an obvious generalization for $\mathbb{Z}_N$ symmetric models
like the sine-Gordon model. From (\ref{eq:anz5},\ref{eq:anz6}) one 
can see that this quantity is closely related to the index of the operator
$AA^\dagger$. For the simple models
considered here, where only one spatial 
direction is nontrivial, $ \theta{'}(k)$  is easily  obtained
from the Dirac equation in the asymptotic
regions $x\to\pm\infty$, far away from the kink \cite{Rebhan:1997iv}.
But as we will see below, in case of a less trivial background
like the vortex, the formulation as surface term
will provide essential simplifications \cite{Wimmer:2003PhD}.

\subsection{Dimensional reduction and evanescent counterterms}

In the above, we have effectively used the 't Hooft-Veltman version
of dimensional regularization \cite{'tHooft:1972fi}
in which the space-time dimensionality $n$
is made larger than the dimension of interest. 
This is possible in a supersymmetric way
if the model of interest can be obtained from a 
higher dimensional supersymmetric 
model by dimensional reduction. 
The nontrivial corrections to the central charge of the kink
come from the ``genuine'' momentum
operator $\tilde P_y$, and are due to 
a spontaneous
breaking of parity.

We now comment on how the central charge anomaly can be recovered
from Siegel's version of dimensional regularization 
\cite{Siegel:1979wq,Siegel:1980qs,Capper:1980ns}
where $n$ is smaller than the dimension of space-time and where one keeps
the number of field components fixed, but lowers the number of
coordinates and momenta from 2 to $n<2$. At the one-loop level one
encounters 2-dimensional $\delta_\mu^\nu$ coming from
Dirac matrices, and $n$-dimensional $\hat\delta_\mu^\nu$ from
loop momenta. An important concept which is going to play a role are
the evanescent counter\-terms \cite{Bonneau:1980zp,Bonneau:1980jx,deWit:1993qv}
involving the factor $\frac{1}{\epsilon}\hat{\hat{\delta}}{}_\mu^\nu
\gamma_\nu\psi$, where $\hat{\hat\delta}{}_\mu^\nu\equiv \delta_\mu^\nu-
\hat\delta_\mu^\nu$ has only $\epsilon=2-n$ nonvanishing components.

The supercurrent is given by 
$j_\mu=-(\not\!\partial\vf+U(\vf))\gamma_\mu\psi$.
In the trivial vacuum, expanding into quantum fields yields
\begin{equation}\label{jmuexp}
j_\mu=-\left(\not\!\partial\eta+U'(v)\,\eta
+
\2 U''(v)\,\eta^2
\right)\gamma_\mu
\psi + \frac{1}{\sqrt{2\lambda}}\delta\mu^2\gamma_\mu\psi.
\end{equation}
Only matrix elements with one external fermion are divergent.
The term involving $U''(v)\eta^2$ in (\ref{jmuexp}) gives rise to
a divergent scalar tadpole that is cancelled 
completely by the counterterm $\delta\mu^2$ (which
itself is due to an $\eta$ and a $\psi$ loop). The only other
divergent diagram is due to the term involving $\not\!\partial\eta$
in (\ref{jmuexp}) and has the form of a $\psi$-selfenergy. Its
singular part reads
\begin{equation}
\<0| j_\mu |p\>^{\rm div} = i U''(v)
\int_0^1 dx \int \frac{d^n \kappa}{(2\pi)^n} 
\frac{\not\!\kappa \gamma_\mu\! \not\!\kappa
}{ [\kappa^2 + p^2 x(1-x) + m^2]^2}u(p).
\end{equation}
Using $\hat\delta_\mu^\nu\equiv \delta_\mu^\nu-\hat{\hat\delta}{}_\mu^\nu$
we find that under the integral
$$\not\!\kappa \gamma_\mu\!\! \not\!\kappa = - \kappa^2(\delta_\mu^\lambda-
\frac{2}{n}\hat \delta_\mu^\lambda)\gamma_\lambda=
\frac{\epsilon}{n}\kappa^2\gamma_\mu - \frac{2}{n}\kappa^2
\hat{\hat\delta}{}_\mu^\lambda \gamma_\lambda$$
so that
\begin{equation}
\<0| j_\mu |p\>^{\rm div} = \frac{U''(v)}{2\pi}\frac{\hat{\hat\delta}{}_\mu^\lambda}{\epsilon} \gamma_\lambda u(p).
\end{equation}
Hence, the regularized one-loop contribution to the susy 
current contains the evanescent operator
\begin{equation}
j_\mu ^{\rm div} = \frac{U''(\vf)}{2\pi}\frac{\hat{\hat\delta}{}_\mu^\lambda
}{\epsilon} \gamma_\lambda  \psi.
\end{equation}
It is called evanescent because the numerator vanishes in
strictly $n=2$; for $n \not= 2$ it has a pole, but in matrix elements this
composite operator gives a finite contribution.
$j_\mu ^{\rm div}$ is by itself a conserved quantity, because all fields
depend only on the $n$-dimensional coordinates, but it
has a nonvanishing contraction with $\gamma^\mu$.
The latter 
gives rise to an anomalous contribution to the 
renormalized conformal-susy current
$\not \!x j_\mu^{\rm ren.}$ where $j_\mu^{\rm ren.}=j_\mu-j_\mu^{div}$,
\begin{equation}
\partial^\mu (\not\!x j_\mu^{\rm ren.})_{\rm anom.}=-
\gamma^\mu j_\mu^{\rm div}= -\frac{U''}{2\pi}\psi.
\end{equation}
(There are also nonvanishing
nonanomalous contributions
to $\partial^\mu (\not\!x j_\mu)$ because our model is not
conformal-susy invariant at the classical level \cite{Fujikawa:2003gi,Fujikawa:2003nm}.)

Ordinary susy on the other hand is unbroken; there is no anomaly
in the divergence of $j_\mu^{\rm ren.}$. A susy variation of
$j_\mu$ involves the energy-momentum tensor and the
topological central-charge current $\zeta_\mu$
according to
\begin{equation}
\delta j_\mu = -2T_\mu{}^\nu \gamma_\nu \epsilon - 2 \zeta_\mu \gamma^5 
\epsilon,
\end{equation}
where classically $\zeta_\mu=\epsilon_{\mu\nu}U
\partial^\nu \varphi$.

At the quantum level,
the counter-term $j_\mu^{\rm ct}=-j_\mu^{\rm div.}$ induces
an additional contribution to the central charge current
\begin{equation}
\zeta_\mu^{\rm anom}=\frac{1}{4\pi}
\frac{\hat{\hat\delta}{}_\mu^\nu}{\epsilon}
\epsilon_{\nu\rho}\partial^\rho U'
\end{equation}
which despite appearances is a {\em finite} quantity: using that total
antisymmetrization of the three lower indices has to vanish
in two dimensions gives
\begin{equation}
\hat{\hat\delta}{}_\mu^\nu
\epsilon_{\nu\rho} = \epsilon \epsilon_{\mu\rho}+
\hat{\hat\delta}{}_\rho^\nu
\epsilon_{\nu\mu} 
\end{equation}
and together with the fact the $U'$ only depends on $n$-dimensional
coordinates this finally yields
\begin{equation}\label{zetaanom}
\zeta_\mu^{\rm anom}=\frac{1}{4\pi} \epsilon_{\mu\rho} \partial^\rho U'
\end{equation}
in agreement with the anomaly in the central charge as obtained
previously.

We emphasize that $\zeta_\mu$ itself does not require the
subtraction of an evanescent counterterm. The latter only
appears in the susy current $j_\mu$, which gives rise
to a conformal-susy anomaly in $\not \!x j_\mu$.
A susy variation of the latter
shows that it forms a conformal current multiplet involving besides
the dilatation current $T_{\mu\nu}x^\nu$ and the Lorentz current
$T_\mu{}^\nu x^\rho\epsilon_{\nu\rho}$
also a current 
$j_{(\nu)}^{(\zeta)\mu}=x^\rho \epsilon_{\rho\nu}\zeta^\mu$.
We identify this with the conformal central-charge current, which
is to be distinguished from the ordinary central-charge current $\zeta_\mu$.

Since $\partial_\mu j_{(\nu)}^{(\zeta)\mu}=\epsilon_{\mu\nu}\zeta^\mu$,
and $\epsilon_{\mu\nu}$ is invertible, the entire
central-charge current $\zeta^\mu$ enters in the divergence
of the conformal central-charge current,
whereas in the case of the conformal-susy current it was 
the contraction $\gamma_\mu j^\mu$.

The current $j^{(\zeta)}$ thus has the curious property of being
completely determined by its own divergence. For this reason it 
is in fact not associated with any continuous symmetry
(as is also the case for the ordinary central-charge current,
which is of topological origin). In the absence
of classical breaking of conformal invariance it is conserved
trivially by its complete disappearance and then there is no
symmetry generating charge operator. Nevertheless, in the 
conformally noninvariant susy kink model this current is nonvanishing
and has in addition to its nonanomalous divergence an 
anomalous one, namely the
anomalous contribution to the central charge current
inherited from the evanescent counterterm in the renormalized susy current.

\subsection{Multiplet shortening, BPS saturation and fermion parity}

We construct representations of the strictly two dimensional algebra 
\begin{equation}
   \label{eq:anzb3}
   Q^{+2}=H+Z\ \ ,\ \ Q^{-2}=H-Z\ \ ,\ \ \{Q^+,Q^-\}=2P,
\end{equation}%
where $Q^{\pm\dagger}=Q^\pm$ are hermitian, by going to the 
rest frame ($P=0$)
in the topological sector, i.e. with non-vanishing central 
charge. $M$ and $Z$ are then ordinary numbers. In the general case, $M\neq |Z|$,
the irreducible representations are two-dimensional 
$\{|\Sigma_-\rangle|,
\Sigma_+ \rangle\}$. The super\-charges can be represented as 
\begin{equation}
 \label{eq:anzb6}
  \hat{Q}^+=\sqrt{M+Z}\ \sigma_1\ \ , \ \ \hat{Q}^-=\sqrt{M-Z}\ \sigma_2. 
\end{equation}

For BPS states the 
absolute value of the central charge is per definition 
equal to  the energy of this state, 
i.e. for the eigen values in (\ref{eq:anzb3}) we have 
\begin{equation}
 \label{eq:anzb7}
        M-|Z|=0.
\end{equation}
We choose $Z=-M$ which corresponds in our convention to the kink. 
The algebra (\ref{eq:anzb3}) in such a BPS representation  becomes now 
\begin{equation}
 \label{eq:anzb8}
        \hat{Q}^{+2}=0\ \ ,\ \ \hat{Q}^{-2}=2M\ \ ,\ \ \{\hat{Q}^+,\hat{Q}^-\}=0.
\end{equation}%
Because of the hermiticity of $\hat{Q}^+$, (\ref{eq:anzb8}) implies   
$||\hat{Q}^+|\Sigma\rangle||^2=0$ and thus 
\begin{equation}
 \label{eq:anzb9}
 \hat{Q}^+|\Sigma\rangle=0.
\end{equation}
This equation is equivalent to  (\ref{eq:anzb7}) for the definition of BPS states 
and means that the BPS states are left invariant by half of the supersymmetry,
namely $Q^+$ in our case. Operators and states can be characterized by the 
cohomology of the operator $Q^+$. Analogous to BRST exact operators
which have vanishing matrix elements for physical states we can say 
that each operator which is $Q^+$-- exact  has vanishing expectation value 
for BPS states:
\begin{equation}
 \label{eq:anzb10}
        \ca{O}=\{Q^+,\ca{O}{'}\} \Rightarrow\langle BPS|\ca{O}|BPS\rangle=0.
\end{equation}%

For the other supercharge in (\ref{eq:anzb8}) $\hat{Q}^-$, which acts nontrivially, 
there exist two 
inequivalent irreducible representations,
\begin{equation}
 \label{eq:anzb11}
        \hat{Q}^-|\Sigma\rangle=\pm\sqrt{2M}|\Sigma\rangle,
\end{equation}
which are connected by a  $\mathbb{Z}_2$ transformation 
$\psi\rightarrow -\psi$ for all fermions, which is clearly a 
symmetry for each fermionic action.
Thus the irreducible representation is one-dimensional 
and the \emph{fermionic} operator
is diagonal \cite{Losev:2000mm}. This is the reason why it was  originally 
thought that
multiplet shortening does not occur in two dimensions 
\cite{Witten:1978mh,Rebhan:1997iv,Shifman:1998zy}. 
Therefore  a reducible two dimensional 
representation for the soliton states was assumed such that the fermion
parity operator $(-1)^F$ is still defined. For a reducible two dimensional
representation $\{|\Sigma_\mathrm{b}\rangle,\ |\Sigma_\mathrm{f}\rangle\}$
we may choose:
\begin{equation}
  \label{eq:anzb12}
  \hat{Q}^-=\sqrt{2M}\ \sigma_1\ \ ,\ \ (-1)^F=\sigma_3,
\end{equation}
so that $(-1)^F$ is diagonal in this representation and $\hat{Q}^-$ has fermion parity
$-1$, i.e. $\{\hat{Q}^-,(-1)^F\}=0$. 
Note that this is the direct sum of the two
inequivalent irreducible representations (\ref{eq:anzb11}), which are 
obtained as $\hal(|\Sigma_\mathrm{b}\rangle\pm|\Sigma_\mathrm{f}\rangle)$.

Witten and Olive \cite{Witten:1978mh} argued that in four dimensional
susy gauge theories the number of particle states is not changed in the 
Higgs phase, although massive representations have  $2^{\ca{N}}$ times  as 
many states than massless
one. Thus, they concluded that the Higgs phase corresponds
to a BPS saturated representation which  has the same number
of physical states as the
massless representation. Because of this \emph{multiplet shortening} the 
BPS saturation should be protected against perturbative corrections
since they cannot change the number of particle states. 

The counting of susy soliton states in two dimensions is 
somewhat peculiar (see below) and the loss of fermion parity (\ref{eq:anzb11})
suggested a two dimensional representation, as for 
the non-susy soliton \cite{Jackiw:1976fn}, and thus no multiplet shortening
would occur.
In \cite{Shifman:1998zy}, nevertheless BPS-saturation was assumed, 
to match  the central charge correction to  the mass correction obtained
in \cite{Nastase:1998sy}. The crucial relation for BPS saturation is 
the annihilation by one super charge (\ref{eq:anzb9}). It was stated
that this relation is protected without multiplet shortening, by analogous 
arguments that constrain supersymmetry breaking \cite{Witten:1982df}. A simple 
argument shows that this is not sufficient. Assume that in some \emph{approximation}
a reducible multiplet is BPS saturated, i.e. $\hat{Q}^+|\Sigma_i\rangle=0$.
Since the operator $Q^+$ is hermitian its representation is of the form
(because of the reducibility the $\mathbb{Z}_2$--grading 
through $(-1)^F$ exists)
\begin{equation}
  \label{eq:anzb13}
  \hat{Q}^+=\begin{pmatrix}0&M\\M^\dagger&0\end{pmatrix},
\end{equation}%
and the BPS states can be separated in zero-eigen states of $M$ and $M^\dagger$.
To answer the question if the multiplet remains BPS saturated under perturbations
(corrections) we consider the quantity%
\footnote{Note that BPS states ($Q^{+2}=0$) contribute 
$\textrm{Tr}|_{BPS}(-1)^F=n_{\textrm{b}}^0-n_{\textrm{f}}^0$, whereas for 
non-BPS states is $Q^{+2}=\langle H+Z\rangle > 0$ such that their 
contribution vanish for $\beta\rightarrow\infty$.}  
\begin{equation}
  \label{eq:anzb14}
  \lim_{\beta\rightarrow\infty}\textrm{Tr}\left[(-1)^F\ e^{-\beta\ Q^{+2}} \right]
  =n_{\textrm{b}}^0-n_{\textrm{f}}^0=\textrm{Ind}(M),
\end{equation}
which is the difference between the number of zero-eigenvalue eigen states 
(singlets) of $M$ 
and $M^\dagger$. That this index is invariant under perturbative 
corrections can be seen quite analogously to the arguments of 
\cite{Witten:1982df}. If a state is no longer annihilated by $Q^+$, 
say for example one with fermionic parity, then also the 
bosonic super-partner state is no longer annihilated:
\begin{equation}
  \label{eq:anzb15}
  0\neq Q^+|\textrm{f}\ \rangle\sim Q^+Q^-|\textrm{b}\rangle=
  -Q^-Q^+|\textrm{b}\rangle \rightarrow Q^+|\textrm{b}\rangle\neq 0.
\end{equation}
So the difference in the number of singlet states is unchanged under 
perturbative corrections, and thus it can be 
calculated in a semi-classical approximation. 
So what can this index tell us? In case that it would be 
nonzero, there would exist, at least one, BPS saturated state, 
which is then protected against quantum corrections. But in this case
$(-1)^F$ is no longer defined as we will see immediately. 
If the index vanishes in an approximation, 
$n_{\textrm{b}}^0-n_{\textrm{f}}^0=0$, the number of fermionic and
bosonic singlets coincides. The trivial case is of course that they both vanish 
already in the approximation and there are no BPS states. In the 
nontrivial case there exist susy pairs of BPS singlets in the approximation, 
but susy does not protect them from being lifted 
pairwise  above the BPS bound as described in (\ref{eq:anzb15}). 
But this is exactly the case of the $\mathbb{Z}_2$ symmetric two-dimensional
multiplet (\ref{eq:anzb12}). So the equality between the mass correction and 
the anomalous contribution to the central charge needs a different explanation.
In fact, it was found that one has to give up the usual fermion parity for the 
topological soliton state which is then a single-state short 
super-multiplet (\ref{eq:anzb11}) \cite{Losev:2000mm}. If we look now again on the
 BPS saturation equation (\ref{eq:anzb9}), we see immediately
that a lift above the BPS bound would give a twice as long irreducible multiplet
(\ref{eq:anzb6}) which cannot be caused by perturbative corrections. 
So, in the absence of other mechanisms as for example a difference in 
a conserved quantum number, 
multiplet shortening is a necessary condition for BPS saturation
being protected.

Up to now we have only discussed  abstract representations of the susy algebra
(\ref{eq:anzb3}). In a quantum field theory the operators in the algebra
(\ref{eq:anzb3}) and their representations correspond to Heisenberg operators
of conserved, i.e. time-independent, charges and Heisenberg states. In general 
neither operators nor states in the Heisenberg picture are known explicitly. 
Instead one quantizes the field operators in the interaction picture
in terms of creation/annihilation operators which are defined w.r.t. a 
perturbative ground state. The canonical commutation relations imply an algebra
for the mode coefficients which usually has to be 
represented in an irreducible manner. 
This determines which kind of the above representations is realized in the
(perturbative) quantum field theory.
From the fermionic creation/annihilation operators (\ref{eq:fermetc}) one obtains an 
infinite dimensional
Clifford algebra of pairs of generators $\gamma_k^+=(b_k+b_k^\dagger)$ and 
$\gamma_k^-=i(b_k-b_k^\dagger)$ and the single generator $b_0$ corresponding to the zero-mode. 
This is 
quite analogous to an odd dimensional Clifford algebra, and the operator
$b_0$ plays the role of the $\gamma_5$ of the even-dimensional 
algebra $\gamma_k^{\pm}$. So there are two inequivalent representations of the 
full algebra governed by the sign of the ``gamma five'' operator $b_0$.
Because $b_0$ has to anti-commute with the $b_k$'s it cannot
be represented as a number, as in the quasi-classical approximation \cite{Losev:2000mm}. 
The $b_k$-algebra can be represented as usual on  a 
Fock space, constructed from the  Clifford vacuum $|\Omega\rangle$ with 
$b_k|\Omega\rangle=0$. The whole algebra, including $b_0$ can then 
be realized by two inequivalent irreducible representations 
\cite{Goldhaber:2000ab}:
\begin{equation}
  \label{eq:ms18}
  |s_\pm\rangle=\hal(1\pm b_0)|\Omega\rangle\ \ ,\ \ 
  b_0 |s_\pm\rangle=\pm |s_\pm\rangle\ \ ,\ \ b_k |s_\pm\rangle=0.
\end{equation}%
According to the usual fermion-parity counting $b_0$ is an odd, i.e. fermionic
operator, and thus the ground states $|s_\pm\rangle$ are half
fermionic and half bosonic. But in two dimensions there is less
distinction between fermions and bosons. In fact,  since there are no
rotations in one spatial dimension, the definition of fermion parity is more
abstract. Indeed, the assignment of fermion number
to different vacua depends on the
sign of the eigenvalue of the fermion mass matrix at the considered vacuum 
\cite{Witten:1982df}. In the case of the kink this means
that if the vacuum with $\varphi=+v$ is defined to be bosonic 
the vacuum  at $-v$ is automatically fermionic. Now a topological state
like the kink connects these two vacua with opposite fermion parity 
which heuristically
explains that this state cannot have a definite fermion parity
in the usual sense.

Let us now check if also semi-classically the BPS saturation condition
is satisfied. With the regularized mode expansion for the quantum fields 
(\ref{eq:bosfluct},\ref{eq:ferm})
and the BPS equation $\partial_x\vf_K+U=0$ one obtains
{\cl{2pt}
\begin{eqnarray}
  \label{eq:ms18a}
  Q^+|s_\pm\rangle&=&\int dx\ [(\dot\eta - \partial_y\eta)
\psi^+ +(\partial_x\eta +U{'}\eta)\psi^-]
  |s_\pm\rangle\nonumber+O(\hbar)\\
  &=&i\int\frac{d^{d-1}\ell}{(2\pi)^{d-1}}\intsum \frac{dk}{2\pi}
    (\sqrt{\omega-\ell}-\sqrt{\omega+\ell})a_k^\dagger b_k^\dagger
   |s_\pm\rangle=0. 
\end{eqnarray}}%
So 
both states are BPS saturated semi-classically. 
\section{Supersymmetric vortices in 2+1 
dimensions}
\label{sec:vor}

Following \cite{Rebhan:2003bu} we next consider the
Abrikosov-Nielsen-Olesen 
\cite{Abrikosov:1957sx,Nielsen:1973cs,deVega:1976mi,Taubes:1980tm}
vortex solution of the abelian Higgs model in 2+1 dimensions
which has a supersymmetric
extension \cite{Schmidt:1992cu,Edelstein:1994bb} 
{\rm (for related models see \cite{Gorsky:1999hk,Vainshtein:2000hu})}
such that
classically the Bogomolnyi bound \cite{Bogomolny:1976de} is saturated.
We implement dimensional regularization of the $2+1$-dimensional
$\mathcal N=2$ vortex by
dimensionally  reducing
the  $\mathcal N=1$ abelian Higgs model in $3+1$ dimensions.
We confirm the results of 
\cite{Schmidt:1992cu,Lee:1995pm,Vassilevich:2003xk}
that in a particular gauge (background-covariant Feynman-'t Hooft)
the sums over zero-point energies of fluctuations in
the vortex background cancel completely, but 
contrary to \cite{Schmidt:1992cu,Lee:1995pm} we find
a nonvanishing quantum correction to the vortex mass
coming from a finite renormalization of the expectation value
of the Higgs field in this gauge \cite{Wimmer:2003PhD,Vassilevich:2003xk}.
In contrast to \cite{Schmidt:1992cu}, where a null result for
the quantum corrections to the central charge was stated,
we show that the central charge receives also a net nonvanishing
quantum correction, namely from a nontrivial phase in the fluctuations
of the Higgs field in the vortex background, which
contributes to the central charge even though the latter is
a surface term that can be evaluated far away from the vortex.
The correction to the central charge exactly matches the
correction to the mass of the vortex. 

In Ref.~\cite{Lee:1995pm}, it was claimed
that the usual
multiplet shortening arguments 
which prove BPS saturation
would not be applicable to the $\mathcal N=2$ vortex since
in the vortex background there would be two
rather than one fermionic
zero modes \cite{Lee:1992yc}, leading to two short multiplets which
have the same number of states as one long multiplet.\footnote{Incidentally,
Refs.\
\cite{Lee:1995pm,Lee:1992yc} considered
the supersymmetric abelian Higgs model augmented by
a Chern-Simons term.}
We show however that the extra zero mode postulated in \cite{Lee:1995pm}
has to be discarded because its gaugino component is singular,
and that only after doing so there is agreement with the
results from index theorems \cite{Weinberg:1981eu,Lee:1992yc,Hori:2000kt}.
For this reason, standard multiplet shortening arguments do
apply, explaining the BPS saturation at the quantum level that we observe
in our explicit one-loop calculations.
















\subsection{The model}

The superspace action for the vortex in terms of 3+1-dimensional
superfields
contains
an $\mathcal N=1$ abelian vector multiplet and an 
$\mathcal N=1$ scalar multiplet,
coupled as usual, together with a Fayet-Iliopoulos term but without
superpotential,
\be
\mathcal L = 
\int d^2 \theta\, W^\alpha W_\alpha
+ \int d^4\theta\, \bar\Phi\, e^{-eV} \Phi + \kappa \int d^4\theta\, V.
\ee
In terms of 2-component spinors in 3+1 
dimensions, the action reads\footnote{\label{fnconv}Our conventions
are $\eta^{\mu\nu}=(-1,+1,+1,+1)$, $\chi^{\alpha}=
\epsilon^{\alpha\beta}\chi_\beta$ and $\5\chi^{\dot\alpha}=
\epsilon^{\dot\alpha\dot\beta}\5\chi_{\dot\beta}$ with
$\epsilon^{\alpha\beta}=\epsilon_{\alpha\beta}=
-\epsilon^{\dot\alpha\dot\beta}=-\epsilon_{\dot\alpha\dot\beta}$
and $\epsilon^{12}=+1$. In particular we have
$\5\psi_{\.\alpha}=
(\psi_\alpha)^*$ but $\5\psi^{\.\alpha}=-
(\psi^\alpha)^*$. Furthermore, 
$\5\sigma^\mu_{\.\alpha\beta}=(-\mathbf 1,\vec\sigma)$
with the usual representation for the Pauli matrices $\vec\sigma$,
and $\5\sigma^{\mu\.\alpha\beta}=\sigma^{\mu\beta\.\alpha}$
with $\sigma^{\mu\alpha\.\beta}=(\mathbf 1,\vec\sigma)$.}
\bea
\mathcal L &=& -\4 F_{\mu\nu}^2
+\5\chi^{\.\alpha} i \5\sigma_{\.\alpha\beta}^\mu
\6_\mu \chi^\beta +\2 D^2+
(\kappa-e|\phi|^2)D\nn
&&-|D_\mu \phi|^2+ \5\psi^{\.\alpha} i \5\sigma_{\.\alpha\beta}^\mu
D_\mu \psi^\beta +|F|^2 +\sqrt2 e \left[ \phi^* \chi_\alpha \psi^\alpha
+\phi \5\chi_{\.\alpha} \5\psi^{\.\alpha} \right],
\eea
where $D_\mu=\6_\mu - ieA_\mu$ when acting on $\phi$ and $\psi$,
and $F_{\mu\nu}=\6_\mu A_\nu-\6_\nu A_\mu$.
Elimination of the auxiliary field $D$ yields the scalar potential
$\mathcal V=\2 D^2=\2 e^2(|\phi|^2-v^2)^2$ with $v^2\equiv {\kappa/e}$.

This model is invariant under the following transformation rules:
\begin{eqnarray}
  \label{eq:swz2}
\delta A_\mu=\epsilon_\alpha\sigma_\mu^{\alpha\dot\alpha}\bar\chi_{\dot\alpha}
  -\bar\epsilon^{\dot\alpha}\bar\sigma_{\mu\dot\alpha\alpha}\chi^\alpha
  \ &,&\ \delta D=
i(\epsilon_\alpha\sigma^{\mu\alpha\dot\alpha}\partial_\mu\bar\chi_{\dot\alpha}+
  \bar\epsilon^{\dot\alpha}\bar\sigma^\mu_{\dot\alpha\alpha}\partial_\mu\chi^\alpha)\\
  \delta\chi^\alpha=-iF_{\mu\nu}\sigma^{\mu\nu\alpha}{}_{\beta}\epsilon^\beta-D\epsilon^\alpha
 \ &,&\ \delta\bar\chi_{\dot\alpha}=
  iF_{\mu\nu}\bar\sigma^{\mu\nu}{}_{\dot\alpha}{}^{\dot\beta}
\bar\epsilon_{\dot\beta}-D\bar\epsilon_{\dot\alpha},
\end{eqnarray}}%
for the gauge multiplet; the matter multiplet transforms as
{\cl{2pt}
\begin{eqnarray}
  \label{eq:swz3}
  \delta\phi&=&-\sqrt{2}\epsilon_\alpha\psi^\alpha\ ,\ \   
  \delta\psi_\alpha=-i\sqrt{2}\bar\epsilon^{\dot\alpha}\bar\sigma^\mu_{\dot\alpha\alpha}D_\mu\phi+
   \sqrt{2}F^* \epsilon_\alpha \nonumber\\
  \delta\phi^*&=&\sqrt{2}\bar\epsilon^{\dot\alpha}\bar\psi_{\dot\alpha}\ ,\ \ \ \  
  \delta\bar\psi_{\dot\alpha}=-i\sqrt{2}(D_\mu\phi)^\ast
\bar\sigma_{\dot\alpha\alpha}^\mu\epsilon^\alpha
   +\sqrt{2}F\bar\epsilon_{\dot\alpha}\nonumber\\
  \delta F&=&-i\sqrt{2}\epsilon_\alpha\sigma^{\mu\alpha\dot\alpha}
D_\mu\bar\psi_{\dot\alpha}+2e\phi^*
    \epsilon_\alpha\chi^\alpha\ ,\ 
  \delta F^*=-i\sqrt{2}\bar\epsilon^{\dot\alpha}\bar\sigma^\mu_{\dot\alpha\alpha}D_\mu\psi^\alpha+2e\phi
    \bar\epsilon^{\dot\alpha}\bar\chi_{\dot\alpha},
\end{eqnarray}}
where
\begin{equation}
  \sigma^{\mu\nu\alpha}{}_{\beta}=
\frac{1}{4}(\sigma^{\mu\alpha\dot\beta}\bar\sigma_{\dot\beta\beta}^\nu-
   \sigma^{\nu\alpha\dot\beta}\bar\sigma_{\dot\beta\beta}^\mu)\quad,\quad
   \bar\sigma^{\mu\nu}{}_{\dot\alpha}{}^{\dot\beta}=
   \frac{1}{4}(\bar\sigma^\mu_{\dot\alpha\beta}\sigma^{\nu\beta\dot\beta}-
                                  \bar\sigma^\nu_{\dot\alpha\beta}\sigma^{\mu\beta\dot\beta}).                             
\end{equation}
As always, before eliminating auxiliary fields,
the transformation rules of the gauge multiplet do not
depend on the matter fields, and the transformation rules themselves
are
gauge covariant and lead to a closed superalgebra.

In 2+1 dimensions, dimensional reduction gives an $\mathcal N=2$ 
model involving,
in the notation of \cite{Lee:1995pm
},
a real scalar $N=A_3$ and two Dirac spinors 
$\psi=(\psi^\alpha)$, $\chi=(\chi^\alpha)$.

Completing squares in the bosonic part of the classical
Hamiltonian density for time-independent fields
one finds the Bogomolnyi equations
and the central charge
\bea
\mathcal H&=&\4 F_{kl}^2+|D_k \phi|^2+\2 e^2 (|\phi|^2-v^2)^2 \nn
&=& \2 |D_k \phi + i\epsilon_{kl} D_l \phi|^2
+\2 \left(F_{12}+e(|\phi|^2-v^2)\right)^2 \nn&&
+\frac{e}{2} v^2 \epsilon_{kl}F_{kl} - i \6_k (\epsilon_{kl}\phi^* D_l \phi)
\quad \mbox{with $k,l=1,2$.}
\eea
The classical central charge reads
\be
Z=\int d^2x \, \epsilon_{kl} \6_k
\left( e v^2 A_l - i \phi^* D_l \phi \right),
\ee
where asymptotically $D_l\phi$ tends to zero exponentially fast.
Classically, BPS saturation $E=|Z|=2\pi v^2 |n|$ holds when
the BPS equations $(D_1 \pm iD_2)\phi \equiv D_\pm \phi=0$
and $F_{12} \pm e(|\phi|^2-v^2)=0$ are satisfied, where the
upper and lower sign corresponds to vortex and antivortex, respectively.
In this paper we use the vortex solution.
For winding number $n$, it is given by
\be
\pv = e^{in\theta} f(r), \quad
eA_+^{\mathrm V} = -i e^{i\theta}\frac{a(r)-n}{r},
\quad A_\pm^{\mathrm V} \equiv A_1^{\mathrm V} \pm i A_2^{\mathrm V}
\ee
or, alternatively,
\be
\pv=\pv^1+i\pv^2=\left( \frac{x^1+ix^2}{r}\right)^n f(r), 
\quad eA_k^{\mathrm V}=\frac{\epsilon_{kl}x^l}{r}\frac{a(r)-n}{r},
\ee
where $f'(r)=\frac{a}{r}f(r)$ and $a'(r)=r e^2(f(r)^2-v^2)$
with boundary conditions \cite{Taubes:1980tm}
\bea
&a(r\to\infty)=0, &f(r\to\infty)=v,\nn
&a(r\to0)=n+O(r^2), &f(r\to0)\to r^n+O(r^{n+2}).
\eea

\subsection{Fluctuation equations}

For the calculation of quantum corrections to a vortex solution we
decompose $\phi$ into a classical background part $\pv$ and a quantum
part $\eta$. Similarly, $A_\mu$ 
is decomposed as $A_\mu^{\mathrm V}+a_\mu$, where
only $A_\mu^{\mathrm V}$ with $\mu=1,2$ is nonvanishing.
We use a background $R_\xi$  
gauge fixing term \cite{'tHooft:1971rn,Fujikawa:1972fe}
which is quadratic in the
quantum fields,
\be\label{Lgfix}
\mathcal L_{\rm g.fix}=
-\frac{1}{2\xi} (\6_\mu a^\mu - ie\xi(\pv \eta^* - \pv^* \eta ))^2.
\ee
The corresponding Faddeev-Popov Lagrangian reads
\be
\mathcal L_{\rm ghost}=
b \left( \6_\mu^2 - e^2 \xi \left\{ 2\,|\pv|^2 + 
\pv \eta^* + \pv^* \eta  \right\} \right) c\,.
\ee

The fluctuation equations in 2+1 dimensions have been
given in \cite{Schmidt:1992cu,Lee:1995pm} for the
choice $\xi=1$ (Feynman-'t Hooft gauge
) which leads 
to important simplifications. We shall mostly use this
gauge choice when considering fluctuations in the solitonic
background, but will carry out renormalization in the
trivial vacuum for general $\xi$ to highlight some of
the gauge dependences.

Because we are going to consider dimensional regularization
by dimensional reduction of the 3+1 dimensional model, we shall
need the form of the fluctuation equations with
derivatives in the $x^3$ direction included.
(This one trivial extra dimension will eventually be turned into
$\epsilon\to0$ dimensions.)

In the 't Hooft-Feynman gauge,
the part of the bosonic action quadratic in the quantum fields
reads ($m=0,1,2$ but $\mu=0,1,2,3$)
\bea
&&\mathcal L_{\rm bos}^{(2)} 
= -\2 (\6_\mu a_m)^2 - \2 (\6_\mu a_3)^2 - e^2 |\pv|^2 a_\mu^2\nn&&\quad
- |D_\mu^{\mathrm V} \eta|^2 - e^2 (3|\pv|^2-v^2) |\eta|^2
-2ie a^\mu \left[\eta^* D_\mu^{\mathrm V} \pv - \eta (D_\mu^{\mathrm V} \pv)^*
\right].\quad
\eea
In the trivial vacuum, which corresponds to $\pv\to v$ and $A_\mu^{\mathrm V}
\to 0$, the last term vanishes, but in the solitonic vacuum it
couples
the linearized field equations for the fluctuations $B\equiv(\eta,a_+/\sqrt2)$
with $a_+=a_1+ia_2$ 
to each other according to ($k=1,2$)
\be\label{Bfluceq}
(\6_3^2-\6_t^2)B=
\left(
\begin{array}{cc}
-(D_k^{\mathrm V})^2+e^2(3|\pv|^2-v^2)  &  i\sqrt2 e (D_- \pv) \\
-i\sqrt2 e (D_- \pv)^* & -\6_k^2+2 e^2 |\pv|^2
\end{array}
\right)B.
\ee
The quartet $(a_3,a_0,b,c)$ with $b,c$ the Faddeev-Popov
ghost fields has diagonal field equations at the linearized level
\be
(\6_\mu^2 - 2 e^2 |\pv|^2) Q=0, \quad Q\equiv(a_3,a_0,b,c).
\ee

For the fermionic fluctuations, which we group as
$U=\binom{ \psi^1 }{ \5\chi^{\.1} }$,
$V=\binom{ \psi^2 }{ \5\chi^{\.2} }$,
the linearized field equations
read
\be\label{UVeqs}
LU=i(\6_t+\6_3) V, \quad L^\dagger V = i(\6_t-\6_3) U,
\ee
with
\be
L=
\left(\begin{array}{cc}
iD_+^{\mathrm V} & \sqrt2 e \pv \\
-\sqrt2 e \pv^*  & i\6_-
\end{array}\right), \quad
L^\dagger =
\left(\begin{array}{cc}
iD_-^{\mathrm V} & -\sqrt2 e \pv \\
\sqrt2 e \pv^*  & i\6_+
\end{array}\right).
\ee

Iteration shows that $U$ satisfies the same second order equations
as the bosonic fluctuations $B$,
\bea
&&L^\dagger L U=(\6_3^2-\6_t^2) U, \quad L^\dagger L B=(\6_3^2-\6_t^2) B\\
&&L L^\dagger V=(\6_3^2-\6_t^2) V,
\eea
with $L^\dagger L$ given by (\ref{Bfluceq}),
whereas $V$ is governed by a diagonal equation with
\be\label{LLd}
LL^\dagger=\left(
\begin{array}{cc}
-(D_k^{\mathrm V})^2+e^2|\pv|^2+e^2 v^2  &  0 \\
0 & -\6_k^2+2 e^2 |\pv|^2
\end{array}
\right).
\ee
(In deriving these fluctuation equations we used the BPS equations throughout.)

\subsection{Renormalization}



At the classical level, the energy and central charge of vortices are
multiples of $2\pi v^2$ with $v^2=\kappa/e$. Renormalization
of $v^2$, even when only by finite amounts,
will therefore contribute directly to the 
quantum mass and central charge of the $\mathcal N=2$ vortex, 
a fact that has been neglected in the original
literature \cite{Schmidt:1992cu,Lee:1995pm}
on quantum corrections to the $\mathcal N=2$ vortex.

In 2+1 dimensions, it is possible, just as in the case of the
1+1-dimensional supersymmetric kink, to adopt a ``minimal''
renormalization scheme where the wave function renormalization
constants are set to unity, and only $v^2$ is renormalized.
The renormalization of $v^2$ is then fixed by the requirement
of vanishing tadpoles in the trivial sector of the 2+1 dimensional
model.
The calculation can be conveniently performed by using dimensional
regularization of the 3+1 dimensional $\mathcal N=1$ model. By going down
in the number of spatial dimensions, $3\to3-(1-\epsilon)$, and
setting $\epsilon\to0$ eventually, we have a supersymmetry preserving
regularization method in analogy to the embedding of the
supersymmetric kink as a domain wall in up to $2+1$ dimensions.
 
For the calculation of
the tadpoles we decompose $\phi=v+\eta\equiv v+(\sigma+i\rho)/\sqrt2$, where
$\sigma$ is the Higgs field and $\rho$ the would-be Goldstone boson.
With the gauge fixing term (\ref{Lgfix}) and $\xi=1$ all fields (including
the Faddeev-Popov fields $b$ and $c$) have the same mass $|m|=\sqrt2 e v$.
This gauge choice also avoids mixed $a_\mu$-$\rho$ propagators,
but there are mixed $\chi$-$\psi$ propagators, which can be
diagonalized\footnote{The kinetic terms for $s$ are $-\5s \gamma^\mu\6_\mu s$
where $\bar s=s^\dagger \sigma_3$ and $\gamma^\mu=\{-i\sigma_3,
i\sigma_3 \sigma^m\}$ with $m=1,2,3$, and likewise for $d$.
Note also that $(s^\dagger)^{\.\alpha}=-\5s^{\.\alpha}$
according to footnote \ref{fnconv}.}
by introducing new spinors $s=(\psi+i\chi)/\sqrt2$
and $d=(\psi-i\chi)/\sqrt2$ with mass terms
$\frac{im}{2}(s_\alpha s^\alpha - d_\alpha d^\alpha)+h.c$.


The part of the interaction Lagrangian which is relevant
for $\sigma$ tadpoles to one-loop order is given by
\be
\mathcal L^{\rm int}_{\sigma-{\rm tadpoles}}=
e(\chi_\alpha \psi^\alpha+\5\chi_{\.\alpha} \5\psi^{\.\alpha})\,\sigma
-\frac{em}{2}(\sigma^2+\rho^2)\,\sigma-em(a_\mu^2+
\xi b\,c- \delta v^2)\,\sigma, 
\ee
where $b$ and $c$ are the Faddeev-Popov fields.   

The one-loop contributions to the $\sigma$ tadpole thus read
\bea\label{sigmatadpoles}
&&\!\!\!\includegraphics[bb=0 400 350 475
]{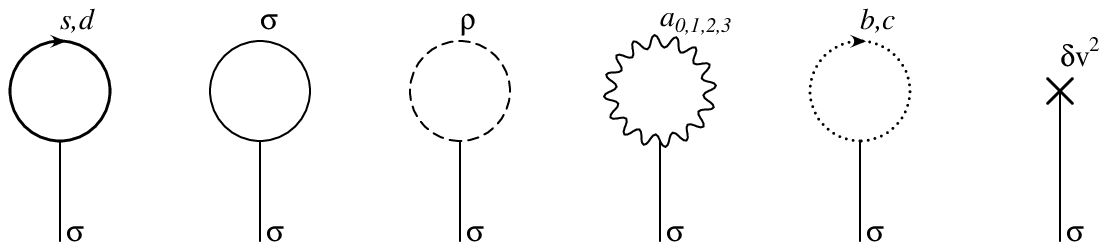}\nn
&& =(-em)\times\\
&&\!\!\!\!\!\!\{-2 {\rm tr}{\mathbf 1_2}I(m) + \frac{3}{2}I(m) + \2 I(\xi^{\frac{1}{2}}m) 
+ [3I(m)+\xi I(\xi^{\2}m)] - \xi I(\xi^{\2}m) -\delta v^2 \},\quad \nonumber
\eea
where 
\be
I(m)=\int \frac{d^{3+\epsilon}k}{(2\pi)^{3+\epsilon}}\frac{-i}{k^2+m^2}
=-\frac{m^{1+\epsilon}}{(4\pi)^{1+\epsilon/2}} 
\frac{\Gamma(-\2-\frac{\epsilon}{2})}{\Gamma(-\2)}=-\frac{m}{4\pi}+O(\epsilon).
\ee
Note that because we use dimensional reduction the component $a_3$
is kept also in the limit of 2+1 dimensions. In fact, the result
equals that of ordinary dimensional regularization in 2+1 dimensions,
where $a_3$ appears as an additional scalar field.

Requiring that the sum of tadpole diagrams (\ref{sigmatadpoles}) vanishes 
fixes $\delta v^2$,
\be\label{deltav2Rxi}
\delta v^2 = \2\left(I(m)+I(\xi^{\2}m)\right)\Big|_{D=3} = 
-\frac{1+\xi^{\2}}{8\pi}m.
\ee
Since in dimensional regularization there are no poles in odd dimensions
at the one-loop level, the result for $\delta v^2$ is finite, but it
is nonvanishing. Because the classical mass of the vortex is
$M_{\mathrm V}=2\pi v^2 = \pi m^2/e^2$, the counterterm $\delta v^2$
is the only one that is of importance to the one-loop corrections
to $M_{\mathrm V}$.

While the latter statement holds generally, the result for $\delta v^2$
does depend on the renormalization scheme adopted. If we consider
more general renormalization schemes, which also allow for
wave function renormalization, the latter do not affect
the mass of the vortex, because that is evaluated at a stationary
point. However, they may affect the definition of $\delta v^2$.
For example, $Z_\phi\not=1$ introduces a further counter-term
to (\ref{sigmatadpoles}) such that the left-hand-side of
(\ref{deltav2Rxi}) is replaced by $\delta v^2 - v^2 \delta Z_\phi$.
While the residue of
the scalar propagator is finite and does not enforce
a nontrivial $\delta Z_\phi$ to dispose of a UV divergence, 
one can nevertheless choose
a finite value of $\delta Z_\phi$ precisely such that $\delta v^2=0$.
By allowing for a nontrivial wave function renormalization
of the vector boson field, $Z_A^{1/2}=Z_e^{-1}$, one can accommodate
further renormalization conditions (for example an
on-shell definition of the elementary masses, residues, 
and coupling constants), which generically involve a nontrivial
$\delta v^2$.
In the following we shall however
stick to the ``minimal'' scheme with $Z_\phi=1$, which has been used
predominantly also in the case of the 1+1 dimensional kink (for
a detailed discussion of other renormalization schemes in the
latter context see \cite{Rebhan:2002uk}).

Since in general
$\delta v^2$ is gauge-parameter dependent, the remaining
contributions to vortex mass and central charge must be gauge dependent, too,
so that the final result is gauge independent when expressed
in terms of physical parameters. (Notice that $v^2$ is not
such a physical parameter, but in any well-defined renormalization
scheme it can be related to the physical mass and coupling constant.)

The gauge breaking term in fact breaks \SUSY, but because the
final result should be gauge-choice independent, we should get
the correct result for the mass and central charge from this $x$-space
\SUSY-breaking approach.
One could also use a superspace approach and fix the U(1) gauge
symmetry without breaking rigid supersymmetry, but in the
presence of solitons the background superspace formalism
leads to some problems, as we sketch
in the appendix.

To gain further insight into the occurrence of tadpoles and their
contribution to the quantum mass of the $\mathcal N=2$ vortex, we briefly
consider an extension of the vortex model in which the $U(1)$ anomaly
in 3+1 dimensions cancels. In the 3+1 dimensional model,
the U(1) coupling of the vector multiplet to the chiral multiplet
is chiral (in terms of 4-component
Majorana spinors it contains the matrix $\gamma_5$),
and in order to cancel the chiral U(1) anomaly, additional scalar multiplets
would be needed such that the sum over charges vanishes, $\sum_i e_i=0$.
As it turns out, this anomaly is also of concern in 2+1 dimensions.
There it does not entail a local gauge anomaly
when $\sum_i e_i\not=0$, but leads to 
a parity-violating Cherns-Simon term
\cite{Redlich:1984kn,Redlich:1984dv,Aharony:1997bx}.

The simplest extension fulfilling this condition is massless super-QED,
consisting of a vector multiplet coupled to two chiral multiplets
with opposite electric charges and 
still without superpotential \cite{Davis:1997bs}.
In such a theory, the Fayet-Iliopoulos term 
is not generated in any order of
perturbation theory if it is not present classically \cite{Fischler:1981zk}.
As we now briefly show, this does not imply the absence
of quantum corrections to $v^2=\kappa/e$ if $v^2$ is nonzero.

In massless super-QED, the kinetic
term, Yukawa coupling and the scalar coupling to $D$ for the new
terms are the same as for $\psi$ and $\phi$, except for an opposite overall
sign. Eliminating $D$, the potential becomes
\be
\mathcal V=\2 e^2 \left(|\phi|^2-|{\widetilde \phi}|^2-
\frac{\kappa}{e}\right)^2
\ee
where $\widetilde\phi$ is the new complex scalar. We locate the vortex solution
again in $\phi$, and use the same gauge fixing term as in (\ref{Lgfix}).%
\footnote{In fact, it has been shown in \cite{Penin:1996si}
that the only solutions with nonvanishing winding number that are
time-independent are the original vortex solutions, with $\widetilde\phi=0$
and the vortex located in $\phi$. An analogous result holds also
for the kink \cite{Striet:2002mb}.}
We find then two new couplings which produce $\sigma$-tadpoles
\be
\mathcal L^{\rm int,extra}_{\sigma-{\rm tadpoles}}=
\2 e m \sigma (\widetilde\sigma^2+\widetilde\rho^2)
\ee
However, these new fields are massless and therefore the additional
tadpole diagrams vanish in dimensional regularization.


On the other hand, expanding about $\widetilde\phi=0$ 
the $\widetilde\sigma$ field does not have tadpole diagrams---there
is only a $\chi\widetilde\psi\widetilde\sigma$ vertex left,
but this does not give rise to a tadpole diagram because
$\widetilde\psi$ does not mix with $\chi$ and $\psi$ (or $s$ and $d$)
when $\widetilde\phi$ vanishes.

Hence, the finite renormalization of $v^2$ in this extension
of the 2+1-dimensional
susy Higgs models is identical to that of the simpler model
we have introduced above, provided the minimal renormalization
scheme with trivial wave function renormalization constants
is employed.\footnote{It should be noted, however, that the
situation is more complicated in 3+1 dimensions, since there
one cannot do without wave function renormalization.}

In the following we shall 
show that the nonzero result for $\delta v^2$
leads to a nonvanishing mass correction for the $\mathcal N=2$ vortex in 2+1 dimensions
and is in fact required to match an equally nonzero correction to the
central charge in order that the BPS bound remains saturated.

\subsection{Quantum corrections to mass and central charge}

The expressions for the central charge and stress tensor can be
constructed from the classical action without any gauge artefacts.
However, when one evaluates one-loop corrections, one uses the
gauge-fixing term to obtain propagators and well-defined
fluctuation equations, and then 
one should use the quantum expression for $H$.
In that case one should
consider sums over zero-point energies
including unphysical degrees of freedom and Faddeev-Popov ghosts.
This can be done in a well-defined manner by using
dimensional regularization by dimensional reduction from
the 3+1 dimensional model. Using this method, the central
charge contains the standard 2+1 dimensional terms and, as a potential
anomalous contribution,
a remainder from the momentum operator in the extra spatial dimension.

\subsubsection{Mass}

At the one-loop level, the quantum mass of a solitonic state is
given by
\be\label{M1loop}
M=M_{\rm cl}+\2\sum \omega_{\rm bos} - \2\sum \omega_{\rm ferm}
+\delta M
\ee
where $M_{\rm cl}$ is the classical mass expressed in terms of
renormalized parameters, $\delta M$ represents the effects
of the counter-terms to these renormalized parameters, and
the sums are over zero-point energies in the soliton background
(the zero-point energies in the trivial vacuum, which one should
subtract in principle, cancel in a \SUSY\ theory).

In the $\xi=1$ gauge
the sum over zero-point energies is formally
\bea\label{zeromodesums}
\2\sum \omega_{\rm bos} - \2\sum \omega_{\rm ferm}
&=& \sum \omega_{\eta} + \sum \omega_{a_+} - \sum \omega_U - \sum \omega_V\nn
&=& \sum \omega_U - \sum \omega_V,
\eea
where the quartet $(a_3,a_0,b,c)$ cancels separately.
(Note that in (\ref{zeromodesums}) all frequencies appear twice because
all fields are complex.)

Using dimensional regularization for models with solitons
as developed in \cite{Rebhan:2002uk}, these sums can be made
well defined by replacing all eigen frequencies $\omega_k$ in 2+1
dimensions by $\omega_{k,\ell}=(\omega_k^2+\ell^2)^{1/2}$ where
$\ell$ are the extra momenta, and integrating
over $\ell$:
\be\label{sumUVdimreg}
\sum \omega_U - \sum \omega_V
=\int \frac{d^2k}{(2\pi)^2} \int \frac{d^\epsilon \ell }{ (2\pi)^\epsilon}
\omega_{k,\ell} \int d^2x \,
[|u_1|^2+|u_2|^2-|v_1|^2-|v_2|^2](\mathbf x;\mathbf k)
\ee
where we have written out only the contributions from the continuous
part of the spectrum, using that $L^\dagger L$ and $LL^\dagger$,
which govern $U$ and $V$, respectively,
are isospectral up to zero modes.
In dimensional regularization,
the zero modes of $L^\dagger L$ and $LL^\dagger$ become massless
modes (with energy $\sqrt{\ell^2}$), but they
can be dropped because the $\ell$-integration
is scaleless, and thus vanishes in dimensional
regularization.

The existing literature \cite{Schmidt:1992cu,Lee:1995pm}
also proves that the spectral densities of the continuous spectrum
is equal for $U$ and $V$, so there is a complete cancellation
of the sums over zero-point energies in the $\xi=1$ gauge.
In analogy to the calculation
performed in eqs.~(\ref{eq:anz6}) and (\ref{eq:anz7}), 
we can verify this result explicitly as follows.

Using that $LU_{\mathbf k}=\omega_k V_{\mathbf k}$
and $L^\dagger V_{\mathbf k}=\omega_k U_{\mathbf k}$, we can
write the difference in the spectral densities appearing
in (\ref{sumUVdimreg}) as
\be
\Delta \rho(\mathbf k)=\int d^2x\, [U_{\mathbf k}^\dagger U_{\mathbf k}-
V_{\mathbf k}^\dagger V_{\mathbf k}]
=\int d^2x\, [U_{\mathbf k}^\dagger U_{\mathbf k}-
\omega_k^{-2} (LU)_{\mathbf k}^\dagger LU_{\mathbf k}]
\ee
and partially integrate to obtain a surface term
of the form
\bea\label{Deltarhok}
\Delta \rho(\mathbf k) &=&i \omega_k^{-2}\int d^2x \{
\6_- [u_1^*(i D_+ u_1 + \sqrt2 e \pv u_2)]
+\6_+[u_2^*(i\6_- u_2 - \sqrt2 e \pv^* u_1)]\}\nonumber\\
&=&i \omega_k^{-2}\lim_{r\to\infty}
\oint d\theta \,[u_1^*(-\6_\theta+in)u_1+u_2^* \partial_\theta u_2] = 0,
\eea
which vanishes because 
cyclindrical waves 
decay like $|u_{1,2}|\sim r^{-1/2}$.

Hence, $\sum \omega_U - \sum \omega_V=0$, and
all that remains is a possible renormalization of $v^2$.
As we discussed, in the minimal renormalization scheme
where all wave functions renormalizations are trivial there
is a finite result for $\delta v^2$, and this leads to
a nonvanishing quantum correction of the vortex mass according to
\be\label{MV}
E
= 2\pi |n| (v^2 + \delta v^2|_{\xi=1}) = 
2\pi |n| (v^2 - \frac{m}{4\pi})
\equiv |n| \left( \frac{ \pi m^2 }{ e^2} - \frac{m}{2} \right),
\ee
where 
$\xi=1$, since in other gauges
the fluctuation equations for the
$B$ fields, i.e.\ $\eta,a_+$, no longer match those of the $U$ fermions.
As always in quantum field theory, the explicit form of the
quantum corrections depends on the definition of the parameters
in the Lagrangian, i.e.\ on the renormalization conditions
employed, which we have chosen in the simplest possible manner.
There are of course renormalization schemes which are more
physical such as on-shell renormalization of the parameters
in the trivial vacuum. In the case of kink and kink domain walls,
an extensive analysis of renormalization schemes other than
the minimal one can be found in \cite{Rebhan:2002uk}.

The above result for the vortex mass in the minimal renormalization
scheme
agrees with \cite{Vassilevich:2003xk}, where however
a careful analysis of boundary conditions in the heat-kernel
approach was needed because the vortex had to be put in a box
to discretize the spectrum. In dimensional regularization one does
not need to put the system in a box, and as a consequence there is
no need to study the contributions from these artificial
boundaries.


In the super-QED model considered
at the end of the previous section 
we have found that the additional (tilde) fields do not
change $\delta v^2$ as given in (\ref{deltav2Rxi}).
Concerning the
additional fluctuation equations for these fields, these are even simpler
than that of the minimal model: The $\widetilde\eta$ field
is governed by
\be
(\6_3^2-\6_t^2)\widetilde\eta
=[-(D_k^{*\mathrm V})^2+e^2(v^2-|\pv|^2)]\widetilde\eta=
-D_-^{*\mathrm V} D_+^{*\mathrm V}\widetilde\eta
\ee
where $D^*$ differs from $D$ only in the sign in front of $e$.
The fermionic tilde field equations read
\be
iD_-^{*\mathrm V} \widetilde\psi^1
=i(\6_t+\6_3) \widetilde\psi^2, \quad
iD_+^{*\mathrm V} \widetilde\psi^2=
i(\6_t-\6_3) \widetilde\psi^1.
\ee
Iteration shows that $\widetilde\psi^2$ has the same field
equation as $\widetilde\eta$, so that in the mode sum we
have
\be
\sum \omega(D_-^{*\mathrm V} D_+^{*\mathrm V})
-\sum \omega(D_+^{*\mathrm V} D_-^{*\mathrm V}).
\ee
The same arguments that prove $\sum \omega_U-\sum \omega_V=
\sum \omega(L^\dagger L) - \sum \omega(L L^\dagger)=0$
can now be repeated for these
simpler operators, and so the above result for the mass correction
to the $\mathcal N=2$ vortex in 2+1 dimensions remains unchanged when
considering the vortex of the anomaly-free super-QED model.

\subsubsection{Central charge}

By starting from the \SUSY\ algebra in 3+1 dimensions
one can derive the central charge in 2+1 dimensions as the
component $T^{03}$ of
\be
T^{\mu\nu}=-\frac{i}{4} {\rm Tr} \,\sigma^{\mu \alpha\dot\alpha}\,
\{ \bar Q_{\dot\alpha},J_\alpha^\nu\}
\ee
where $J_\alpha^\nu$ is the \SUSY\ Noether current.

The antisymmetric part of $T^{\mu\nu}$ gives the
standard expression for the central charge density, while
the symmetric part is a genuine momentum in the extra
dimension:
\be
\< Z \> = \int d^2 x \< T^{03} \> 
= \< \tilde Z + \tilde P_3 \>.
\ee
(A similar decomposition is valid for the kink \cite{Rebhan:2002yw}.)

$\tilde Z$ corresponds to the classical expression for the central
charge. Being a surface term, its quantum corrections can be
evaluated at infinity:
\be
\langle \tilde Z \rangle =
\int d^2x \6_k \epsilon_{kl} \langle \tilde\zeta_l \rangle
= \int_0^{2\pi}\!\!\! d\theta \langle \tilde\zeta_\theta \rangle|_{r \to \infty}
\ee
with 
$\tilde\zeta_l=e v^2_0 A_l-i\phi^\dagger D_l \phi$ and $v_0^2=v^2+\delta v^2$.

Expanding in quantum fields $\phi=\pv+\eta$, 
$A=A^{\rm V}+a$ and using that
the classical fields $\pv\to v e^{in\theta}$, $A_\theta^{\rm V}\to n/e$,
$D_\theta^{\rm V} \pv \to 0$ as $r\to \infty$,
we 
obtain to one-loop order 
\bea\label{Zab}
&&\langle \tilde Z \rangle 
= 2\pi n v^2_0 - i \int_0^{2\pi}\!\!\! d\theta \left\langle 
(\pv^*+\eta^\dagger) (D_\theta^{\rm V}-ie a_\theta)(\pv+\eta)
\right\rangle|_{r \to \infty} \nn
&=& 2\pi n \{v^2_0 - \langle \eta^\dagger \eta \rangle |_{r \to \infty}\}
- i \int_0^{2\pi}\!\!\! d\theta \left\{
\left\langle 
\eta^\dagger \6_\theta \eta 
\right\rangle
-ie \pv^* \left\langle a_\theta \eta \right\rangle
-ie \pv \left\langle a_\theta \eta^\dagger \right\rangle
\right\}|_{r \to \infty}\nn
&\equiv& Z_a+Z_b
\eea
where we have used 
$\langle \eta(r\!\to\!\infty) \rangle \to 0$ (which determines $\delta v^2$
at infinity, but at finite $r$, the vacuum expectation
value does not vanish \cite{Shifman:1998zy,Goldhaber:2001rp}),
$\langle a_\theta \rangle = 0$, 
and
$\langle \eta^\dagger \eta a_\theta \rangle = O(\hbar^2)$.

The first contribution, $Z_a$, can be easily evaluated for
arbitrary gauge parameter $\xi$, yielding
\bea
Z_a&=&2\pi n \{v^2_0
-\2(
\langle \sigma\sigma \rangle + \langle \rho\rho \rangle) |_{r \to \infty}\}
\nn&=&
2\pi n \{v^2_0-\2[I(m)+I(\xi^{\2}m)]\}\nn&=&2\pi n (v_0^2-\delta v^2)=
2\pi n v^2.
\eea
If this was all, the quantum corrections to $Z$
would cancel, just as in the naive
calculation of $Z$ in the \SUSY\ kink \cite{Imbimbo:1984nq,Rebhan:1997iv}.

The second contribution in (\ref{Zab}), however, does not
vanish when taking the limit $r\to\infty$. This contribution
is simplest in the $\xi=1$ gauge, where the $\eta$ and $a_\theta$
fluctuations are governed by the fluctuation equations
(\ref{Bfluceq}). In the limit $r\to\infty$ one has
$|\pv|\to v$ and $D_-\pv\to 0$ exponentially. This
eliminates the contributions from $\left\langle a_\theta \eta \right\rangle$
(note that at finite $r$ there is a cross-term in the kinetic
terms for $a_\theta$ and $\eta$).
However, $D_k^2$, which governs the $\eta$ fluctuations,
contains long-range contributions from the vector
potential. Making a separation of variables in $r$ and $\theta$
one finds that asymptotically 
\be
|D_k^{\mathrm V} \eta|^2 \to |\6_r \eta|^2+\frac{1}{r^2} |(\6_\theta-in)\eta|^2
\ee
so that the $\eta$ fluctuations
have an extra phase factor $e^{in\theta}$ compared to those in the
trivial vacuum. We thus have, in the $\xi=1$ gauge,
\bea
Z_b&=&
- i \int_0^{2\pi}\!\!\! d\theta \left\{
\left\langle 
\eta^\dagger \6_\theta \eta 
\right\rangle
-ie \pv^* \left\langle a_\theta \eta \right\rangle
-ie \pv \left\langle a_\theta \eta^\dagger \right\rangle
\right\}|_{r \to \infty}\nn
&=&- i \int_0^{2\pi}\!\!\! d\theta 
\left\langle \eta^\dagger \6_\theta \eta 
\right\rangle_{\xi=1} 
= 2\pi n \left\langle \eta^\dagger \eta \right\rangle_{\xi=1,r \to \infty}
= 2\pi n\, \delta v^2\Big|_{\xi=1}\,.
\eea
This is exactly the result for the one-loop correction to the
mass of the vortex in
eq.~(\ref{MV}), implying saturation of the BPS bound 
provided that there are now no anomalous contributions to the
central charge operator 
as there are in the case in the $\mathcal N=1$ \SUSY\
kink \cite{Rebhan:2002yw}.

In 
dimensional regularization by dimensional
reduction from a higher-dimensional model such anomalous
contributions to the central charge operator come from
a finite remainder of the extra momentum operator \cite{Rebhan:2002yw}. 
The latter is
given by \cite{Lee:1995pm}
\bea
Z_c=\< \tilde P_3\>&\!=\!&
\int d^{2}x \< F_{0i}F_{3i}
+ (D_0 \phi)^\dagger D_3 \phi
+ (D_3 \phi)^\dagger D_0 \phi \right. \nn
&&\qquad\qquad\left.
- i \bar \chi \bar\sigma_0 \6_3 \chi
-i \bar \psi \bar\sigma_0 D_3 \psi \>.
\eea
Inserting mode expansions for the quantum fields one
immediately finds that the bosonic contributions vanish 
because of symmetry in the extra trivial dimension.
However, this is not the case for the fermionic fields,
which have a mode expansion of the form
\be
\genfrac{(}{)}{0pt}{0}{U}{V} =
\int\frac{d^\epsilon \ell }{ (2\pi)^{\epsilon/2}}
{\Sigma}\!\!\!\!\!\!\int_k
\frac{1}{\sqrt{2\omega}} \Bigl\{
b_{k,\ell}\, e^{-i(\omega t-\ell z)}
\left(\genfrac{}{}{0pt}{0}{
  \genfrac{}{}{0pt}{}{\sqrt{\omega-\ell}\, u_1}{ \sqrt{\omega-\ell}\, u_2}
}{
  \genfrac{}{}{0pt}{}{\sqrt{\omega+\ell}\,v_1 }{  \sqrt{\omega+\ell}\,v_2}
}\right)
+d_{k,\ell}^\dagger \times(c.c.)
\Bigr\},
\ee
where we have not written out explicitly the zero-modes
(for which $\omega^2=\ell^2$).
The fermionic contribution to $Z_c$ reads 
\bea
Z_c&=&\< \tilde P_3\>=\nn
&=& \int\frac{d^\epsilon \ell }{ (2\pi)^\epsilon}
{\Sigma}\!\!\!\!\!\!\int_k \frac{\ell^2}{2\omega}
\int d^2x \left[ |u_1|^2+|u_2|^2-|v_1|^2-|v_2|^2 \right](\mathbf x;\mathbf k)
\eea
where $\omega=\sqrt{\omega_k+\ell^2}$, so that the $\ell$ integral
is nontrivial in dimensional regularization. 
Only the continuous spectrum can contribute because
zero modes give scaleless integrals
which vanish in dimensional regularization and if there were
other discrete states, they would cancel between $U$ and $V$.
However, the $x$-integration
over the mode functions $u_{1,2}$ and $v_{1,2}$
produces their spectral densities, and we
find
\be
Z_c=\int \frac{d^2k}{(2\pi)^2}\frac{d^\epsilon \ell }{ (2\pi)^\epsilon}
\frac{\ell^2}{2\omega}\Delta\rho(\mathbf k) =0
\ee
because $\Delta\rho(\mathbf k) =0$ as we have seen
in (\ref{Deltarhok}).
Hence, $|Z|=|Z_a+Z_b|=E$, so that the BPS bound is saturated at
the (one-loop) quantum level.

\subsection{Fermionic zero modes and multiplet shortening}

Massive representations of the Poincar\'e supersymmetry algebra 
for which the absolute value
of the central charge equals the energy, i.e.~when the BPS bound is
saturated, contain as many states as 
massless representations, which is half of that of
massive representations for which the BPS bound is not saturated. 
These results
also apply in 2+1 dimensions for the $\mathcal N=2$ super-Poincar\'e algebra
\cite{Lee:1995pm}.

A particular multiplet of states is obtained by taking the vortex solution,
and acting on it with the \SUSY\ generators of the $\mathcal N=2$ \SUSY\ algebra,
which contains two complex charges $Q^+$, $Q^-$, and their hermitian
conjugates $(Q^+)^\dagger$ and $(Q^-)^\dagger$. One of these charges,
$Q^+$, annihilates the vortex solution, while the other one, $Q^-$, is to
linear order in quantum operators proportional to the annihilation
operator of a fermionic zero mode. 

However, if there indeed is a second fermionic
zero mode in the model as claimed in \cite{Lee:1995pm}\footnote{\rm
In the literature one can in fact find two different conventions for
indicating the number of fermionic zero modes.
Like Refs.~\cite{Lee:1995pm,Hori:2000kt} we only count the number of zero
modes in the fermionic quartet $(U,V)$ and not additionally
those in the corresponding conjugated fields $(U^\dagger,V^\dagger)$.
One zero mode in this way of counting then corresponds to
a {\em pair} of creation/annihilation operators. Alternatively
one may count the zero modes in both $(U,V)$ and $(U^\dagger,V^\dagger)$
and thus ascribe one zero mode
to each creation or annihilation operator.
The latter way of counting is
employed for instance in Ref.~\cite{Vainshtein:2000hu}.}, 
in second quantization it would be present in the
mode expansion of the fermionic quartet $U$ and $V$,
\be
\genfrac{(}{)}{0pt}{0}{U}{V} 
=a_{\mathrm I} 
\left(\genfrac{}{}{0pt}{0}{
  \genfrac{}{}{0pt}{}{\psi_{\mathrm I}^1 }{ \bar\chi^{\dot 1}_{\mathrm I}} 
}{
 \genfrac{}{}{0pt}{}{0 }{ 0} }\right)
+a_{\mathrm{I\!I}} 
\left(\genfrac{}{}{0pt}{0}{
  \genfrac{}{}{0pt}{}{\psi_{\mathrm{I\!I}}^1 }{ \bar\chi^{\dot 1}_{\mathrm{I\!I}}} 
}{
 \genfrac{}{}{0pt}{}{0 }{ 0} }\right)
+\mbox{non-zero modes.}
\ee
As a result, there would then be a quartet of BPS states
\be\label{quartet}
\ket{v},\;
a_{\mathrm I}^\dagger\ket{v},\;
a_{\mathrm{I\!I}}^\dagger\ket{v},\;
a_{\mathrm I}^\dagger a_{\mathrm{I\!I}}^\dagger\ket{v}
\ee
comprising two short multiplets of $\mathcal N=2$ \SUSY, which
are degenerate and together have as many states as one
long multiplet without BPS saturation.  As stressed in \cite{Lee:1995pm},
the standard argument
for stability of BPS saturation under quantum corrections
from multiplet shortening \cite{Witten:1978mh} thus would not be applicable.

However, we shall now show that there is in fact only a
single fermionic zero mode in a vortex background with
winding number $n=1$ \cite{Rebhan:2003bu}.
To this end,
we first observe that the zero modes must lie
in $U$, because $V$ is governed by the operator $LL^\dagger$ of
Eq.~(\ref{LLd}), whose only zero mode solution is $V_0\equiv 0$.
A zero mode for $U$ must satisfy $LU=0$, 
and to analyse this equation we follow \cite{Lee:1995pm} and
set $\psi^1(x,y)=-i e^{i(j-\2+n)\theta}
u(r)$ and $\bar\chi^{\dot 1}=e^{i(j+\2)\theta}d(r)$.
The equation $LU=0$ reduces then to
\be\label{udeqs}
\left(\begin{array}{cc}
\6_r-({a+j-\2})/{r} & \sqrt2 e f \\
\sqrt2 e f & \6_r+({j+\2})/{r}
\end{array}\right)
\genfrac{(}{)}{0pt}{0}{u}{d}= 0,
\ee
where $f=f(r)$ and $a=a(r)$ satisfy $f'=\frac{a}{r}f$ and $a'=re^2(f^2-v^2)$.
Iterating this equation yields
\be\label{uzm}
\left( \6_r^2 + \frac{1}{r}\6_r - \frac{(j-\2)^2}{r^2} - 2 e^2 f^2 \right)
\frac{u}{f} = 0.
\ee
Given a solution for $u$, the corresponding solution for $d$ follows from
$LU=0$.

For given $j$, this equation has two independent solutions, a linear
combination of which yields solutions which decrease exponentially
fast as $r\to\infty$. Hence, {\em both} solutions should be regular
at $r=0$. For $j\not=\2$, one has, using $f(r\to0)\sim r^n$,
\be
\psi^1 \sim u\sim r^n(C_1 r^{j-\2} + C_2 r^{-(j-\2)}) \quad
\mbox{for $r\to 0$}
\ee
which selects for $n=1$ only $j=-\2$. This solution is the zero mode
that is  obtained by acting with
$Q^-$ on the background solutions,
which gives $\psi^1=-i D_-\phi_{\rm V}/\sqrt2=-i\sqrt2 f'$,
$\bar\chi^{\dot1}=F_{12}=-e(f^2-v^2)$. For $j=\2$, one finds for $n=1$ near
$r=0$
\be\label{seczmorig}
\psi^1 \sim C_1\,(x+iy)+C_2\,(x+iy)\ln r\;.
\ee
For large $r$, $\psi_1 \sim e^{-mr}e^{i\theta}$, as follows from
(\ref{uzm}).
This solution corresponds to the second fermionic zero mode
postulated in Ref.~\cite{Lee:1995pm}.

However, while (\ref{seczmorig}) is finite at the origin,
the associated gaugino component is not: (\ref{udeqs}) implies
that
\be
\bar\chi^{\dot 1}\sim C_2 \frac{e^{i\theta}}{r},
\ee
so this solution has to be discarded when $C_2\not=0$.

Similarly, one can show that for winding number $n>1$ regularity
of the gaugino component generically requires that $j\le-\2$ so that
the correct quantization condition for normalizable fermionic zero modes
is $-n+\2 \le j \le -\2$. Hence, there are $n$ independent
fermionic zero modes, not $2n$ as concluded in \cite{Lee:1995pm}.
It is in fact only the former value that agrees with the
results \cite{Lee:1992yc,Hori:2000kt} 
obtained from the index theorem 
\cite{Weinberg:1981eu}.

{\rm
As has been proved rigorously in \cite{Weinberg:1979er},
in the bosonic sector there are $2n$ zero modes,
which are related to the above
$n$ independent fermionic zero modes by supersymmetry.
In the $R_{\xi=1}$ background gauge 
$\6_\mu a^\mu - ie(\pv \eta^* - \pv^* \eta )=0$,
the bosonic zero modes 
satisfy a set of equations completely equivalent
to those for the fermionic zero modes \cite{Lee:1992yc}.
But
the linearly dependent solutions $\binom{U}{0}$ and $i\binom{U}{0}$
correspond to linearly independent solutions for the bosonic
zero modes $a$ and $\eta$.\footnote{\rm For an analogous case 
see eq.\ (3.8) of 
Ref.~\cite{Weinberg:1979ma}.} 
In particular, for $n=1$, the
$j=-\2$ solution
$(\psi^1=-iu(r),\bar\chi^{\dot 1}=d(r))$ with real $u(r)$ and $d(r)$
corresponds to the bosonic zero mode $\eta(r)=-iu(r)$,
$(a_1,a_2)=(\sqrt2 d(r),0)$, while multiplying the fermionic
solution by $i$ corresponds to the bosonic zero mode
$\eta(r)=u(r)$, $(a_1,a_2)=(0,\sqrt2 d(r))$, which is evidently
linearly independent of the former. 
For both solutions the $R_{\xi=1}$ gauge condition is satisfied due to
the lower component of the field equation (\ref{udeqs}).
Conversely, one can start from the classical vortex solution
and find two independent bosonic zero modes by considering
their derivatives with respect to the $x$ and $y$ coordinates.
Performing a gauge transformation to satisfy the
$R_{\xi=1}$ gauge condition leads one back to the above solutions.
This additionally confirms that the above analysis has
identified all fermionic zero modes in the quartet $(U,V)$.
}

We thus conclude that for the basic vortex (winding number $n=1$)
there is exactly one fermionic zero mode ({\rm corresponding to
one pair of fermionic creation/annihilation operators})
and this gives rise
to a single short multiplet at the quantum level.
Standard multiplet shortening arguments therefore do apply and
explain the preservation of BPS saturation that we verified at
one-loop order.
 

\bigskip
\centerline{Acknowledgments}

This work was supported in part by the National Science Foundation,
 Grant nos. PHY-0140192 and
PHY-0098527, and the Austrian Science Foundation FWF,
project no. 15449.


\appendix

\section{$R_\xi$ gauge in superspace}


It seems natural to use superfields in the calculation of quantum
corrections, since in this way we manifestly preserve rigid \SUSY.
Since we want to descirbe quantum fluctuations about a nontrivial
background, we use the background formalism for superspace. The
action reads $\bar\Phi e^V \Phi$, where $\Phi=\phi_{\rm V}(x)+
\psi(x,\theta)$ and $V=V_{\rm V}(x)+w(x,\theta)$ for abelian
gauge theories. 
Expanding in terms of quantum fields, the
classical action contains again off-diagonal terms quadratic
in quantum fields,
\be
\mathcal L_{\rm kin.} = \int d^4\theta \left[
\phi_{\rm V}^* w \psi + \bar\psi w \phi_{\rm V} +\ldots \right].
\ee
To cancel these, we try to modify the usual $D=4$ gauge fixing
term $D^2 V \bar D^2 V$ into a superspace $R_\xi$ gauge fixing term
\be
\mathcal L_{\rm fix}=-
(\bar D^2w + \bar D^2\frac{1}{\Box}\bar\psi \pv)
(D^2w + D^2\frac{1}{\Box}\psi \pv^*).
\ee
In a trivial (constant) vacuum with $\phi_{\rm V}=v$, the terms
$\bar\psi \phi_{\rm V}$ and $\phi_{\rm V}^* \psi$ are
antichiral and chiral, respectively. So to maintain these chirality
properties we extend $\phi_{\rm V}(x)$ in the soliton sector to
an {\it anti-}chiral superfield $\phi_{\rm V}(x,\theta)=\phi_{\rm V}(x^\mu-
i\bar\theta\bar\sigma^\mu\theta)$. Then $\phi_{\rm V}^*$ is chiral,
and one has the identities
\be
\5D^2 D^2 \frac{1}{\Box} (\psi \phi_{\rm V}^*)=\psi \phi_{\rm V}^*;\qquad
D^2 \5D^2 \frac{1}{\Box} (\5\psi \phi_{\rm V})=\5\psi \phi_{\rm V}.
\ee
The terms $-\int d^4\theta w(\psi \phi_{\rm V}^*+\5\psi \phi_{\rm V})$
in the gauge-fixing term cancel the classical off-diagonal kinetic
terms just as in $x$-space.

However, one is left with the following diagonal terms in the
gauge-fixing term
\be
\int d^4\theta \left[\5D^2 \frac{1}{\Box} (\5\psi \phi_{\rm V}) \right]
\left[D^2 \frac{1}{\Box} (\psi \phi_{\rm V}^*) \right]
= \int d^4\theta  \,\5\psi \phi_{\rm V}\,\frac{1}{\Box}  (\psi \phi_{\rm V}^*)= D^2 \left[ (\5 D^2 (\5\psi \phi_{\rm V})) \frac{1}{\Box} 
(\psi \phi_{\rm V}^*) \right].
\ee
Although $D^2\5D^2 (\5\psi \phi_{\rm V})$ produces a factor $\Box$
which cancels the nonlocal $1/\Box$, further terms with one
or two $D_\alpha$ acting on $\psi \phi_{\rm V}^*$ yield
nonlocal interactions. Thus it seems that one cannot construct
a local $R_\xi$ gauge with $\xi=1$ in superspace.\footnote{
For a recent application of the superspace background formalism
to the kink see Refs.~\cite{Fujikawa:2003gi,Shizuya:2003vm}; 
in this case there are of course no problems with gauge fixing.}


\end{document}